\documentclass[letterpaper]{article}
\usepackage{aaai2026} %
\usepackage{times} %
\usepackage{helvet} %
\usepackage{courier} %
\usepackage[hyphens]{url} %
\usepackage{graphicx} %
\usepackage{natbib} %
\usepackage{caption} %
\usepackage{algorithm}
\usepackage{algorithmic}

\usepackage{amsmath}
\usepackage{amsfonts}
\usepackage{amsthm}
\usepackage{mathrsfs}
\usepackage{mathtools}
\usepackage{adjustbox}
\usepackage{booktabs}
\usepackage{listings}
\usepackage[acronym]{glossaries}
\usepackage{xspace}
\usepackage{circledsteps}
\usepackage{tikz}
\usetikzlibrary{arrows.meta, calc, positioning}
\tikzset{>=latex} %
\usepackage{pgfplots} %
\usepgfplotslibrary{fillbetween} %
\pgfmathdeclarefunction{gauss}{3}{%
    \pgfmathparse{1/(#3*sqrt(2*pi))*exp(-((#1-#2)^2)/(2*#3^2))}%
}
\usepackage[outline]{contour} %
\usetikzlibrary{patterns}
\pgfplotsset{compat=1.12} %

\usepackage{environ}
\usepackage{import}
\usepackage{pgf}
\usepackage{comment}
\DeclareCaptionStyle{ruled}{labelfont=normalfont,labelsep=colon,strut=off}
\usepackage{xcolor}

\usepackage{graphicx}
\usepackage[outdir=./]{epstopdf}
\graphicspath{{./figures/}} %

\usepackage{pifont}%
\newcommand{\cmark}{\ding{51}} %
\newcommand{\xmark}{\ding{55}} %
\newcommand{\xmarkred}{{\color{magenta}\xmark}} %
\definecolor{accent}{RGB}{160,32,240}
\definecolor{accent2}{RGB}{0,141,225}

\usetikzlibrary{shapes.geometric, arrows.meta, positioning, fit, shadows} %

\newcommand{\hp}{hyperparameters\xspace}

\newcommand{\yaml}{\textit{.yaml}\xspace}
\newcommand{\json}{\textit{.json}\xspace}

\newcommand{\cdotx}{\,\cdot\,} %
\newcommand{\data}{\boldsymbol{\mathcal{D}}} %
\renewcommand{\S}{\boldsymbol{\mathcal{S}}} %
\newcommand{\T}{^\top}
\newcommand{\R}{\mathbb{R}}

\newcommand{\N}{\mathbb{N}}
\newcommand{\X}{\mathbb{X}}

\renewcommand{\mid}{\,|\,}
\newcommand{\E}{\mathbb{E}}

\newcommand{\w}{\mathbf{w}} %

\DeclareMathOperator{\diag}{diag}

\newcommand{\A}{\mathbb{A}}
\newcommand{\B}{\boldsymbol{B}}
\newcommand{\Bmax}{\hat{\B}} %
\newcommand{\Bmin}{\check{\B}} %
\newcommand{\norm}[1]{\left|\left|#1\right|\right|}

\newcommand{\W}{\mathbb{W}}

\newcommand{\fossil}{\textsc{Fossil}\xspace}
\newcommand{\lucid}{\textsc{Lucid}\xspace}
\newcommand{\pylucid}{\textsc{PyLucid}\xspace}
\newcommand{\npinterval}{\textsc{npinterval}\xspace}

\newcommand{\trust}{\textsc{TRUST}\xspace}
\newcommand{\gurobi}{\textsc{Gurobi}\xspace}
\newcommand{\alglib}{\textsc{Alglib}\xspace}
\newcommand{\highs}{\textsc{HiGHS}\xspace}
\newcommand{\dreal}{\textsc{dReal}\xspace}
\newcommand{\omnisafe}{\textsc{OmniSafe}\xspace}
\newcommand{\linear}{\texttt{Linear}\xspace}
\newcommand{\barrII}{\texttt{Barr\textsubscript{2}}\xspace}
\newcommand{\barrIII}{\texttt{Barr\textsubscript{3}}\xspace}
\newcommand{\overtaking}{\texttt{Over}\xspace}

\newcommand{\estimator}{\texttt{Estimator}\xspace}

\theoremstyle{definition}

\theoremstyle{plain}

\theoremstyle{remark}

\lstdefinelanguage{iPython}{
basicstyle={\footnotesize\ttfamily},%
numbers=left,numberstyle=\footnotesize,xleftmargin=2em,%
aboveskip=2pt,
belowskip=2pt,
showstringspaces=false,
tabsize=2,
breaklines=true,
morekeywords={as,access,and,break,class,continue,def,del,elif,else,except,exec,finally,for,from,global,if,import,in,is,lambda,not,or,pass,print,raise,return,try,while},%
morekeywords=[2]{abs,all,any,basestring,bin,bool,bytearray,callable,chr,classmethod,cmp,compile,complex,delattr,dict,dir,divmod,enumerate,eval,execfile,file,filter,float,format,frozenset,getattr,globals,hasattr,hash,help,hex,id,input,int,isinstance,issubclass,iter,len,list,locals,long,map,max,memoryview,min,next,object,oct,open,ord,pow,property,range,raw_input,reduce,reload,repr,reversed,round,set,setattr,slice,sorted,staticmethod,str,sum,super,tuple,type,unichr,unicode,vars,xrange,zip,apply,buffer,coerce,intern},%
sensitive=true,%
morecomment=[l]\#,%
morestring=[b]',%
morestring=[b]",%
morestring=[s]{'''}{'''},%
morestring=[s]{"""}{"""},%
morestring=[s]{r'}{'},%
morestring=[s]{r"}{"},%
morestring=[s]{r'''}{'''},%
morestring=[s]{r"""}{"""},%
morestring=[s]{u'}{'},%
morestring=[s]{u"}{"},%
morestring=[s]{u'''}{'''},%
morestring=[s]{u"""}{"""},%
literate=
    *{+}{{{\color{ipython_purple}+}}}1
{-}{{{\color{ipython_purple}-}}}1
{*}{{{\color{ipython_purple}$^\ast$}}}1
{/}{{{\color{ipython_purple}/}}}1
{^}{{{\color{ipython_purple}\^{}}}}1
{?}{{{\color{ipython_purple}?}}}1
{!}{{{\color{ipython_purple}!}}}1
{\%}{{{\color{ipython_purple}\%}}}1
{<}{{{\color{ipython_purple}<}}}1
{>}{{{\color{ipython_purple}>}}}1
{|}{{{\color{ipython_purple}|}}}1
{\&}{{{\color{ipython_purple}\&}}}1
{~}{{{\color{ipython_purple}~}}}1
{==}{{{\color{ipython_purple}==}}}2
{<=}{{{\color{ipython_purple}<=}}}2
{>=}{{{\color{ipython_purple}>=}}}2
{+=}{{{+=}}}2
{-=}{{{-=}}}2
{*=}{{{$^\ast$=}}}2
{/=}{{{/=}}}2,
literate=
    {á}{{\'a}}1 {é}{{\'e}}1 {í}{{\'i}}1 {ó}{{\'o}}1 {ú}{{\'u}}1
{Á}{{\'A}}1 {É}{{\'E}}1 {Í}{{\'I}}1 {Ó}{{\'O}}1 {Ú}{{\'U}}1
{à}{{\`a}}1 {è}{{\`e}}1 {ì}{{\`i}}1 {ò}{{\`o}}1 {ù}{{\`u}}1
{À}{{\`A}}1 {È}{{\'E}}1 {Ì}{{\`I}}1 {Ò}{{\`O}}1 {Ù}{{\`U}}1
{ä}{{\"a}}1 {ë}{{\"e}}1 {ï}{{\"i}}1 {ö}{{\"o}}1 {ü}{{\"u}}1
{Ä}{{\"A}}1 {Ë}{{\"E}}1 {Ï}{{\"I}}1 {Ö}{{\"O}}1 {Ü}{{\"U}}1
{â}{{\^a}}1 {ê}{{\^e}}1 {î}{{\^i}}1 {ô}{{\^o}}1 {û}{{\^u}}1
{Â}{{\^A}}1 {Ê}{{\^E}}1 {Î}{{\^I}}1 {Ô}{{\^O}}1 {Û}{{\^U}}1
{œ}{{\oe}}1 {Œ}{{\OE}}1 {æ}{{\ae}}1 {Æ}{{\AE}}1 {ß}{{\ss}}1
{ç}{{\c c}}1 {Ç}{{\c C}}1 {ø}{{\o}}1 {å}{{\r a}}1 {Å}{{\r A}}1
{€}{{\EUR}}1 {£}{{\pounds}}1,
commentstyle=\color{ipython_cyan}\ttfamily,
stringstyle=\color{ipython_red}\ttfamily,
keepspaces=true,
showspaces=false,
rulecolor=\color{ipython_frame},
numberstyle=\scriptsize\color{halfgray},
backgroundcolor=\color{ipython_bg},
keywordstyle=\color{ipython_green}\ttfamily,
}
\lstdefinelanguage{yaml}{
numbers=none,
xleftmargin=0em,
keywords = {true,false,null,y,n},
sensitive=false,
comment=[l]{\#},
morecomment=[s]{/*}{*/},
commentstyle=\color{commentGreen}\ttfamily,
stringstyle=\color{stringRed}\ttfamily,
keywordstyle=\color{keywordBlue}\bfseries,
moredelim=**[il][\color{commentGreen}{:}\color{keywordBlue}]{:},
morestring=[b]",
morestring=[b]',
literate =  {---}{{\ProcessThreeDashes}}3
{>}{{\textcolor{red}\textgreater}}1
{|}{{\textcolor{red}\textbar}}1
{\ -\ }{{\mdseries\ -\ }}3,
}
\lstdefinelanguage{json}{
numbers=none,
xleftmargin=0em,
keywords = {true,false,null,y,n},
sensitive=false,
stringstyle=\color{stringRed}\ttfamily,
morestring=[b]",
morestring=[b]',
literate=
*{0}{{{\color{keywordBlue}0}}}{1}
{1}{{{\color{keywordBlue}1}}}{1}
{2}{{{\color{keywordBlue}2}}}{1}
{3}{{{\color{keywordBlue}3}}}{1}
{4}{{{\color{keywordBlue}4}}}{1}
{5}{{{\color{keywordBlue}5}}}{1}
{6}{{{\color{keywordBlue}6}}}{1}
{7}{{{\color{keywordBlue}7}}}{1}
{8}{{{\color{keywordBlue}8}}}{1}
{9}{{{\color{keywordBlue}9}}}{1}
}

\definecolor{commentGreen}{RGB}{0, 100, 0}
\definecolor{stringRed}{RGB}{163, 21, 21}
\definecolor{keywordBlue}{RGB}{0, 0, 255}

\definecolor{halfgray}{gray}{0.55}
\definecolor{ipython_frame}{RGB}{207, 207, 207}
\definecolor{ipython_bg}{RGB}{247, 247, 247}
\definecolor{ipython_red}{RGB}{186, 33, 33}
\definecolor{ipython_green}{RGB}{0, 128, 0}
\definecolor{ipython_cyan}{RGB}{64, 128, 128}
\definecolor{ipython_purple}{RGB}{170, 34, 255}

\usepackage{newfloat}
\usepackage{listings}
\DeclareCaptionStyle{ruled}{labelfont=normalfont,labelsep=colon,strut=off} %
\floatstyle{ruled}
\newfloat{listing}{tb}{lst}{}
\floatname{listing}{Listing}
\lstdefinestyle{nonumbers}{numbers=none,numberstyle=\footnotesize,xleftmargin=0em}
\lstdefinestyle{spacednonumbers}{numbers=none,numberstyle=\footnotesize,xleftmargin=0em,aboveskip=15pt,belowskip=15pt,backgroundcolor=\color{white}}
\lstdefinestyle{bgnonumbers}{numbers=none,numberstyle=\footnotesize,xleftmargin=0em,backgroundcolor=\color{white},backgroundcolor=\color{ipython_bg},}

\glsdisablehyper
\newglossaryentry{oop}{
    name={Object-Oriented Programming},
    first={Object-Oriented Programming (OOP)},
    text={OOP},
    description={
            Object-Oriented Programming is a programming paradigm based on the concept of objects, which can hold a state in the form of fields, and methods, in the form of function with access to the object's state.
        }
}
\newglossaryentry{api}{
    name = {Application Programming Interface},
    first = {Application Programming Interface (API)},
    text = {API},
    plural={APIs},
    firstplural={Application Programming Interfaces (APIs)},
    description = {
            An application programming interface (API) defines a way for two software components to communicate with each other.
        }
}
\newglossaryentry{smt}
{
    name={Satisfiability Modulo Theories},
    first={Satisfiability Modulo Theories (SMT)},
    text={SMT},
    description={
            A decision problem for logical formulas with respect to combinations of background theories
            expressed in classical first-order logic with equality.
        }
}
\newglossaryentry{fol}{
    name={First-order logic},
    first={first-order logic (FOL)},
    text={fol},
    description={
            Expressive formal system that allows the following statements:
            \begin{itemize}
                \item Constants ($x, y$)
                \item Functions ($f(x)$)
                \item Relationships ($x > y$)
                \item Connectives ($\land, \lor, \lnot, \implies, \iff$)
                \item Quantifiers ($\forall, \exists$)
            \end{itemize}
        }
}
\newglossaryentry{smtp}{
    name={SMT problem},
    text={SMT problem},
    plural={SMT problems},
    description={
            Decision problems that are true or false depending on whether a given \gls{fol} formula
            is true or false in respect to a given background theory. \\
            In the context of computer science, the background theories are usually include
            arithmetic, bitvectors, arrays, etc.
        }
}
\newglossaryentry{sat}{
    first={Satisfiability (SAT)},
    name={Satisfiability},
    text={SAT},
    description={
            A formula is satisfiable if it is possible to find a set of values for its variables
            that makes the formula true.
            On the other hand, a formula is unsatisfiable if such set does not exists.
            \begin{multline*}
                \text{Example} \\
                (a \lor c) \land (b \lor c) \land (\neg a \lor \neg c) \\
                \text{is satisfiable with the assignment } a = 1, b = 1, c = 0 \\
            \end{multline*}
        }
}
\newglossaryentry{convopt}{
    name={Convex optimization},
    text={convex optimization},
    description={
            A convex optimization problem is a problem where the objective function and the constraints
            are convex functions. \\
            A convex function is a function whose domain is a convex set and that satisfies the following
            property:
            \begin{equation*}
                f(\lambda x + (1 - \lambda)y) \leq \lambda f(x) + (1 - \lambda)f(y)
            \end{equation*}
            for all $x, y \in \text{dom}(f)$ and $\lambda \in [0, 1]$.
        }
}
\newglossaryentry{cnf}{
    first={Conjunctive Normal Form (CNF)},
    name={Conjunctive Normal Form},
    text={CNF},
    description={
            A formula is in \gls{cnf} if it is a conjunction of clauses, where a clause is a
            disjunction of literals. \\
            A literal can be a boolean variable or its negation. \\
            \\
            \begin{equation*}
                \bigwedge_{i=1}^n \bigvee_{j=1}^{m_i} l_{ij}
            \end{equation*}
        }
}
\newglossaryentry{lp}{
    first={Linear Program (LP)},
    name={Linear Program},
    text={LP},
    description={
            A Linear Program is an optimization problem where the objective function and the constraints
            are linear functions. \\
            A typical Linear Program has the following form:
            \begin{equation*}
                \begin{aligned}
                     & \text{maximize}   & c^T x      \\
                     & \text{subject to} & A x \leq b \\
                     &                   & x \geq 0
                \end{aligned}
            \end{equation*}
            where $x \in \mathbb{R}^d$ is the vector of variables to be determined, $c \in \mathbb{R}^d$ and $b \in \mathbb{R}^n$ are vectors of coefficients, and $A \in \mathbb{R}^{n \times d}$ is a matrix of coefficients.
        }
}
\newglossaryentry{qf-lia}{
    name={QF\_LIA},
    text={QF\_LIA},
    description={
            Quantifier-free linear integer arithmetic. In essence, Boolean combinations of inequations between linear polynomials over integer variables.
        }
}
\newglossaryentry{qf-lra}{
    name={QF\_LRA},
    text={QF\_LRA},
    description={
            Quantifier-free linear real arithmetic. In essence, Boolean combinations of inequations between linear polynomials over real variables.
        }
}
\newglossaryentry{nnf}{
    name={Negation Normal Form},
    text={NNF},
    first={Negation Normal Form (NNF)},
    description={
            A formula where the negation operator $\neg$ is only applied to variables. \\
            $\neg a$ is a NNF formula, while $\neg (a \lor b)$ is not.
        }
}
\newglossaryentry{dpll}{
    name={DPLL algorithm},
    text={DPLL},
    first={Davis-Putnam-Logemann-Loveland (DPLL)},
    description={
            A complete \gls{sat} backtracking algorithm for deciding the satisfiability of propositional logic formulae in \gls{cnf}.
        }
}
\newglossaryentry{dp}{
    name={DP algorithm},
    text={DP},
    first={Davis-Putnam (DP)},
    description={
            A complete \gls{sat} algorithms that uses the resolution inference rule to decide the satisfiability of propositional logic formulae in \gls{cnf}.
        }
}
\newglossaryentry{mps}{
    name={Mathematical Programming System},
    text={MPS},
    first={Mathematical Programming System (MPS)},
    description={
            A common file format for representing \gls{lp} problems.
            Developed by IBM, it is supported by most commercial \gls{lp} solvers.
        }
}
\newglossaryentry{soi}{
    name={Sum of Infeasibilities},
    text={SoI},
    first={Sum of Infeasibilities (SoI)},
    description={
            A metric used to evaluate the quality of an \gls{lp} relaxation.
            It is the sum of the infeasibilities of the constraints that are violated by the solution.
        }
}
\newglossaryentry{onnx}{
    name={Open Neural Network Exchange},
    text={ONNX},
    first={Open Neural Network Exchange (ONNX)},
    description={
            An open-source format for representing deep learning models.
            It is supported by a wide range of tools and libraries.
        }
}
\newglossaryentry{cdlc}{
    name={Conflict Driven Learning Clause},
    text={CDLC},
    first={Conflict Driven Learning Clause (CDLC)},
    description={
            A clause that is added to the \gls{cnf} formula after a conflict is detected during the \gls{sat} solving process.
            It is used to prevent the solver from exploring the same path that led to the conflict.
        }
}
\newglossaryentry{rkhs}{
    name={Reproducing Kernel Hilbert Space},
    text={RKHS},
    first={Reproducing Kernel Hilbert Space (RKHS)},
    firstplural={Reproducing Kernel Hilbert Spaces (RKHS)},
    description={}
}
\newglossaryentry{cbc}{
    name={Control Barrier Certificate},
    text={CBC},
    plural={CBCs},
    first={\emph{Control Barrier Certificate} (CBC)},
    firstplural={\emph{Control Barrier Certificates} (CBCs)},
    description={}
}
\newglossaryentry{cbf}{
    name={Control Barrier Certificate},
    text={CBF},
    plural={CBFs},
    first={\emph{Control Barrier Function} (CBF)},
    firstplural={\emph{Control Barrier Functions} (CBFs)},
    description={}
}
\newglossaryentry{gui}{
    name={Graphical User Interface},
    text={GUI},
    first={\emph{Graphical User Interface} (GUI)},
    description={}
}
\newglossaryentry{cli}{
    name={Command Line Interface},
    text={CLI},
    first={\emph{Command Line Interface} (CLI)},
    description={}
}
\newglossaryentry{nn}{
    name={Neural Network},
    text={NN},
    first={Neural Network (NN)},
    firstplural={Neural Networks (NNs)},
    description={}
}
\newglossaryentry{cme}{
    name={Conditional Mean Embedding},
    text={CME},
    plural={CMEs},
    first={\emph{Conditional Mean Embedding} (CME)},
    firstplural={\emph{Conditional Mean Embeddings} (CMEs)},
    description={}
}
\newglossaryentry{rl}{
    name={Reinforcement Learning},
    text={RL},
    first={Reinforcement Learning (RL)},
    description={}
}
\newglossaryentry{mdp}{
    name={Markov Decision Process},
    text={MDP},
    plural={MDPs},
    first={Markov Decision Process (MDP)},
    firstplural={Markov Decision Processes (MDPs)},
    description={}
}

\makeatletter
\pgfdeclareshape{document}{
    \inheritsavedanchors[from=rectangle] %
    \inheritanchorborder[from=rectangle]
    \inheritanchor[from=rectangle]{center}
    \inheritanchor[from=rectangle]{north}
    \inheritanchor[from=rectangle]{south}
    \inheritanchor[from=rectangle]{west}
    \inheritanchor[from=rectangle]{east}
    \backgroundpath{%
        \southwest \pgf@xa=\pgf@x \pgf@ya=\pgf@y
        \northeast \pgf@xb=\pgf@x \pgf@yb=\pgf@y
        \pgf@xc=\pgf@xb \advance\pgf@xc by-10pt %
        \pgf@yc=\pgf@yb \advance\pgf@yc by-10pt
        \pgfpathmoveto{\pgfpoint{\pgf@xa}{\pgf@ya}}
        \pgfpathlineto{\pgfpoint{\pgf@xa}{\pgf@yb}}
        \pgfpathlineto{\pgfpoint{\pgf@xc}{\pgf@yb}}
        \pgfpathlineto{\pgfpoint{\pgf@xb}{\pgf@yc}}
        \pgfpathlineto{\pgfpoint{\pgf@xb}{\pgf@ya}}
        \pgfpathclose
        \pgfpathmoveto{\pgfpoint{\pgf@xc}{\pgf@yb}}
        \pgfpathlineto{\pgfpoint{\pgf@xc}{\pgf@yc}}
        \pgfpathlineto{\pgfpoint{\pgf@xb}{\pgf@yc}}
        \pgfpathlineto{\pgfpoint{\pgf@xc}{\pgf@yc}}
    }
}
\makeatother

\def\extendedversion{1} %

\if\extendedversion1
    \usepackage{tikz}
\usetikzlibrary{shadows,calc}

\def\shadowshift{3pt,-3pt}
\def\shadowradius{0pt}

\colorlet{innercolor}{black!10}
\colorlet{outercolor}{gray!0}

\newcommand\drawshadow[1]{
    \begin{pgfonlayer}{shadow}
        \shade[outercolor,inner color=innercolor,outer color=outercolor] ($(#1.south west)+(\shadowshift)+(\shadowradius/2,\shadowradius/2)$) circle (\shadowradius);
        \shade[outercolor,inner color=innercolor,outer color=outercolor] ($(#1.north west)+(\shadowshift)+(\shadowradius/2,-\shadowradius/2)$) circle (\shadowradius);
        \shade[outercolor,inner color=innercolor,outer color=outercolor] ($(#1.south east)+(\shadowshift)+(-\shadowradius/2,\shadowradius/2)$) circle (\shadowradius);
        \shade[outercolor,inner color=innercolor,outer color=outercolor] ($(#1.north east)+(\shadowshift)+(-\shadowradius/2,-\shadowradius/2)$) circle (\shadowradius);
        \shade[top color=innercolor,bottom color=outercolor] ($(#1.south west)+(\shadowshift)+(\shadowradius/2,-\shadowradius/2)$) rectangle ($(#1.south east)+(\shadowshift)+(-\shadowradius/2,\shadowradius/2)$);
        \shade[left color=innercolor,right color=outercolor] ($(#1.south east)+(\shadowshift)+(-\shadowradius/2,\shadowradius/2)$) rectangle ($(#1.north east)+(\shadowshift)+(\shadowradius/2,-\shadowradius/2)$);
        \shade[bottom color=innercolor,top color=outercolor] ($(#1.north west)+(\shadowshift)+(\shadowradius/2,-\shadowradius/2)$) rectangle ($(#1.north east)+(\shadowshift)+(-\shadowradius/2,\shadowradius/2)$);
        \shade[outercolor,right color=innercolor,left color=outercolor] ($(#1.south west)+(\shadowshift)+(-\shadowradius/2,\shadowradius/2)$) rectangle ($(#1.north west)+(\shadowshift)+(\shadowradius/2,-\shadowradius/2)$);
        \filldraw ($(#1.south west)+(\shadowshift)+(\shadowradius/2,\shadowradius/2)$) rectangle ($(#1.north east)+(\shadowshift)-(\shadowradius/2,\shadowradius/2)$);
    \end{pgfonlayer}
}

\pgfdeclarelayer{shadow} 
\pgfsetlayers{shadow,main}

\newsavebox\mybox
\newlength\mylen

\newcommand\shadowimage[2][]{
\setbox0=\hbox{\includegraphics[#1]{#2}}
\setlength\mylen{\wd0}
\ifnum\mylen<\ht0
\setlength\mylen{\ht0}
\fi
\divide \mylen by 50
\def\shadowshift{0,0}
\def\shadowradius{\the\dimexpr\mylen+\mylen+\mylen\relax}
\begin{tikzpicture}
\begin{scope}
    \clip [rounded corners=\shadowradius * 0.2] (0,0) rectangle coordinate (centerpoint) (\the\wd0, \the\ht0);
    \node[anchor=south west,inner sep=0] (image) at (0,0) {\includegraphics[#1]{#2}};
\end{scope}
\drawshadow{image}
\end{tikzpicture}
}

\fi

\title{LUCID: Learning-Enabled Uncertainty-Aware Certification of Stochastic Dynamical Systems
\if\extendedversion1
(Extended Version)
\fi}

\author {
    Ernesto Casablanca\textsuperscript{\rm 1},
    Oliver Sch\"on\textsuperscript{\rm 1},
    Paolo Zuliani\textsuperscript{\rm 2},
    Sadegh Soudjani\textsuperscript{\rm 3}
}
\affiliations {
    \textsuperscript{\rm 1}Newcastle University, Newcastle upon Tyne, United Kingdom\\
    \textsuperscript{\rm 2}Universit\`{a} di Roma ``La Sapienza'', Rome, Italy\\
    \textsuperscript{\rm 3}Max Planck Institute for Software Systems, Kaiserslautern, Germany\\
    e.casablanca2@ncl.ac.uk, o.schoen2@ncl.ac.uk, zuliani@di.uniroma1.it, sadegh@mpi-sws.org
}

\begin{document}

\maketitle

\begin{abstract}
    Ensuring the safety of AI-enabled systems, particularly in high-stakes domains such as autonomous driving and healthcare, has become increasingly critical. Traditional formal verification tools fall short when faced with systems that embed both opaque, black-box AI components and complex stochastic dynamics. To address these challenges, we introduce \lucid (Learning-enabled Uncertainty-aware Certification of stochastIc Dynamical systems), a verification engine for certifying safety of black-box stochastic dynamical systems from a finite dataset of random state transitions. As such, \lucid is the first known tool capable of establishing quantified safety guarantees for such systems. Thanks to its modular architecture and extensive documentation, \lucid is designed for easy extensibility.

    \lucid employs a data-driven methodology rooted in control barrier certificates, which are learned directly from system transition data, to ensure formal safety guarantees. We use conditional mean embeddings to embed data into a \gls{rkhs}, where an \gls{rkhs} ambiguity set is constructed that can be inflated to robustify the result to out-of-distribution behavior.

    A key innovation within \lucid is its use of a finite Fourier kernel expansion to reformulate a semi-infinite non-convex optimization problem into a tractable linear program. The resulting spectral barrier allows us to leverage the fast Fourier transform to generate the relaxed problem efficiently, offering a scalable yet distributionally robust framework for verifying safety. \lucid thus offers a robust and efficient verification framework, able to handle the complexities of modern black-box systems while providing formal guarantees of safety. These unique capabilities are demonstrated on challenging benchmarks.
\end{abstract}

\begin{links}
    \link{Source Code}{https://github.com/TendTo/lucid}
    \link{Documentation}{https://tendto.github.io/lucid/}
\end{links}

\section{Introduction}
Embodied forms of AI are on the rise, including applications such as autonomous vehicles, robotics, personalized healthcare, and smart infrastructure.
Their core functionality is built around the advances of deep learning, enabling systems to understand and reason in complex and human-like ways to produce a desired behavior.
Whilst the black-box nature of AI components has been a crucial contributor to the widespread success of deep learning, in response to a rapidly evolving legal scrutinization~\cite{veale2021demystifying}, the resulting lack of traceability and certifiability has put a premature halt to its deployment in safety-critical applications.

As deep learning agents are left to choose their actions autonomously in closed-loop interaction with the physical world, they are confronted with a world riddled with randomness and uncertainty.
Efforts to establishing trust in their safe operation have been largely focused on developing tools to verify the input--output behavior of \glspl{nn} embedded in the systems~\cite{liu2021algorithms}.
Unfortunately, there do not exist any tools that could take these results and certify the safety of the entire system, as uncertainty-aware models of closed-loop systems are rarely available and simulators are often too complicated and opaque to perform verification directly~\cite{wongpiromsarn2023formal}. This motivates a holistic black-box treatment~\cite{corso2021survey}, where safety guarantees are to be established from behavioral data and in account of the unknown laws of randomness themselves.

There exist no tools capable of quantifying safety guarantees for complex stochastic closed-loop systems from data.
For the setting where a model of the closed-loop system is available, \npinterval by \citet{harapanahalli2023forward} verifies safety of nonlinear systems with non-deterministic disturbance based on existing \gls{nn} verification tools. See the references therein and the annual friendly competition by \citet{abate2024arch} for adjacent work.
Few data-driven approaches for stochastic systems exist.
\omnisafe is a comprehensive platform for the development of safe \gls{rl} algorithms~\cite{ji2024omnisafe}. However, safe \gls{rl} does generally only \emph{encourage} safer behavior, without providing any rigorous or quantified guarantees on the probability of the absence of unsafe behavior.
As a popular working principle for certifying safety, \glspl{cbc}~\cite{prajna2006barrier} are at the basis of tools such as \trust~\cite{gardner2025trust}, which supports only polynomial dynamics, and \fossil~\cite{edwards2024fossil}, which is model-based, and both being restricted to deterministic systems.

To close this gap (see Table~\ref{tab:qualitative_comparison}), we introduce \lucid, the first verification engine for black-box stochastic dynamical systems with quantified guarantees.
At its core, \lucid learns \glspl{cbc} for unknown systems based solely on data, by constructing an uncertainty-aware estimator of the expected system behavior based on \glspl{cme}~\cite{muandet2017kernel}.
Crucially, the underlying learning framework establishes quantified safety probabilities with distributionally robust guarantees for arbitrary smooth dynamics, thus relaxing the need for restrictive structural assumptions common to related approaches.
For instance, \citet{schoen2024distributionally} learn \glspl{cbc} for systems with polynomial dynamics and \citet{chen2025distributionally} assume systems with known deterministic component.
\begin{table}
    \centering
    \begin{tabular}{ccccccc}
        \toprule
        \textbf{Tool}                                    & \multicolumn{6}{c}{\textbf{Supported Features}}                                                                                                                                                                  \\[.4em]
                                                         & \rotatebox{90}{Guarantees}                      & \rotatebox{90}{Data Driven} & \rotatebox{90}{Stochastic Dyn.}  & \rotatebox{90}{Non-Poly. Dyn.} & \rotatebox{90}{Stat. Correct.} & \rotatebox{90}{Closed Loop} \\
        \midrule
        \textbf{\lucid} (this work)                      & \cmark                                          & \cmark                      & \cmark                           & \cmark                         & \cmark                         & \cmark                      \\
        \trust~\shortcite{gardner2025trust}              & \cmark                                          & \cmark                      & \xmarkred                        & \xmarkred                      & \cmark\footnotemark[1]         & \cmark                      \\
        \fossil~\shortcite{edwards2024fossil}            & \cmark                                          & \xmarkred                   & \xmarkred                        & \cmark                         & NA                             & \cmark                      \\
        \omnisafe~\shortcite{ji2024omnisafe}             & \xmarkred                                       & \cmark                      & \cmark                           & \cmark                         & \xmarkred                      & \cmark                      \\
        \npinterval~\shortcite{harapanahalli2023forward} & \cmark                                          & \xmarkred                   & \cmark/\xmarkred\footnotemark[2] & \cmark                         & NA                             & \cmark                      \\
        \bottomrule
    \end{tabular}
    \caption{Qualitative comparison with existing tools based on their supported features: quantified safety guarantees, data driven, stochastic dynamics, non-polynomial dynamics, statistical correctness guarantees of the learned model (only applicable if data driven), and support for closed-loop systems.}
    \label{tab:qualitative_comparison}
\end{table}
\footnotetext[1]{Assuming the data satisfies persistence of excitation.} %
\footnotetext[2]{Accepts non-deterministic bounded disturbances.} %

Alternative approaches often leave the statistical correctness with respect to the underlying data-generating process unaddressed: \citet{kazemi2020fullLTL} use model-free reinforcement learning, relying on known Lipschitz constants; \citet{lew2021sampling} compute reachable sets for stochastic systems using a sampling-based scheme, which yields only asymptotic guarantees; and \citet{salamati2024data} use a scenario-based method, based on Lipschitz constants and exponentially large datasets.
For data-driven safety verification of stochastic systems via conformal prediction~\cite{lindemann2024formal}, no tools are available.

We summarize the main contributions of this work:
\begin{itemize}
    \item We introduce \lucid, a novel verification engine that learns control barrier certificates from data using kernel-based \glspl{cme}. \lucid uses a tractable reformulation of the problem via a finite Fourier expansion, enabling efficient barrier synthesis based on a linear program.
    \item A robust and extensible software implementation, with both C++ and Python interfaces, supporting configuration via YAML, JSON, or Python scripts, and offering both a \gls{cli} and a \gls{gui}.
    \item Numerical evaluation on a suite of benchmarks, demonstrating \lucid{}’s unique capabilities, robustness, and practical utility.
\end{itemize}

\paragraph{Organization}
The paper is organized as follows.
Section~\ref{sec:theory} provides the theoretical foundations of \lucid, including the safety of black-box dynamical systems, control barrier certificates, conditional mean embeddings, and derivation of the relaxed linear program.
Section~\ref{sec:tool} describes the architecture and functionalities of \lucid alongside a running example, presenting \lucid's components and how they interact.
Section~\ref{sec:experiments} evaluates \lucid on a suite of benchmarks. Finally, Section~\ref{sec:conclusion_and_future_extensions} provides a conclusion and future extensions.

\section{Theoretical Working Principles}\label{sec:theory}
\subsection{Safety of Black-Box Dynamical Systems}

\paragraph{System Description}
Many AI-driven systems exhibit complex, nonlinear, and stochastic behavior that can be modeled as discrete-time stochastic processes with Markovian dynamics:
\begin{equation}
    \label{eq:model}
    \S\colon\quad x_{t+1}  = f(x_t,a_t,w_t),\quad w_t\sim p_w,
\end{equation}
where $f\colon\X\times\A\times\W\rightarrow\X$ is a continuous vector field describing the evolution of the system state $x_t\in\X\subset\R^n$ over time $t\in\smash{\N_{\geq0}}$, driven by control actions $a_t\in\A\subset\R^m$ and process noise $w_t\in\W\subset\R^l$.
The noise is assumed to be drawn from a stochastic distribution $p_w$ in an independent and identically distributed (i.i.d.) manner.
This general formulation subsumes a wide range of systems, including discrete-time \glspl{mdp}.

\begin{figure}
    \centering
    \begin{tikzpicture}
        \foreach \i in {4, ..., 1}{
                \node[name=xa\i] at (\i/4*9,0) {$x_\i$};
            }
        \foreach \i in {4, ..., 1}{
                \node[name=aa\i] at (\i/4*9+.5,0) {$u_{\i}$};
            }
        \foreach \i in {4, ..., 1}{
                \path[draw,->] (xa\i) -- +(0,.5)-| node[left,above,xshift = -0.3cm]{\footnotesize$\pi$} (aa\i);
            }
        \foreach \i in {3, ..., 1}{
                \pgfmathtruncatemacro{\j}{\i + 1}
                \path[draw,->] (aa\i)-- node[above]{\footnotesize$\S$} (xa\j);
            }
    \end{tikzpicture}
    \caption{Evolution of the closed-loop system.}
    \label{fig:closed_loop_system}
\end{figure}
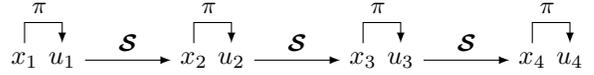

\paragraph{Black-Box Policies}
Here, the focus is on systems $\S$ driven by black-box control policies of the form $\pi\colon\R^n\rightarrow\R^m$, i.e., at every time step $t=0,1,2,\ldots$ a continuous action $a_t\in\A$ is selected based on the current state $x_t$. Such policies could be, for example, neural networks trained via \gls{rl}, or any other black-box function generating actions in $\R^m$ (see Figure~\ref{fig:closed_loop_system}).
The resulting closed-loop system is denoted as $\S^\pi$.

\paragraph{Dataset}
Due to their complexity and opacity, such systems must often be treated as black boxes in their entirety, assuming only access to a finite amount, $N$, of system observations
\begin{equation}
    \data_N\colon \quad \{(x^i,a^i,x^i_+)\}_{i=1}^N,\label{eq:data}
\end{equation}
where every sampled transition from $x^i\in\X$ to a successor $x^i_+\in\X$ is generated as a realization
\begin{equation*}
    (x^i,a^i,x_+^i)\!\sim\!\!\int\! \delta_{f(x^i,a^i,w)}(dx_+^i)\,p_w(dw)\,\mathcal{U}_\X(dx^i)\,\mathcal{U}_\A(da^i),
\end{equation*}
with $\mathcal{U}_\X$ and $\mathcal{U}_\A$ uniform distributions on $\X$ and $\A$, respectively, and $\delta$ indicating the Dirac delta distribution capturing the system dynamics.

Assuming i.i.d. data of the form \eqref{eq:data} and full observability offers statistical guarantees on the consistency of the data-driven CME constructed in Section~\ref{sec:dd_estimation_via_cme}. We point to literature addressing dependent data, assuming either ergodicity, burn-in time, or reduced sample effectiveness w.r.t. the mixing time~\cite{zhang2024guarantees,ziemann2023nonasymptotic}.

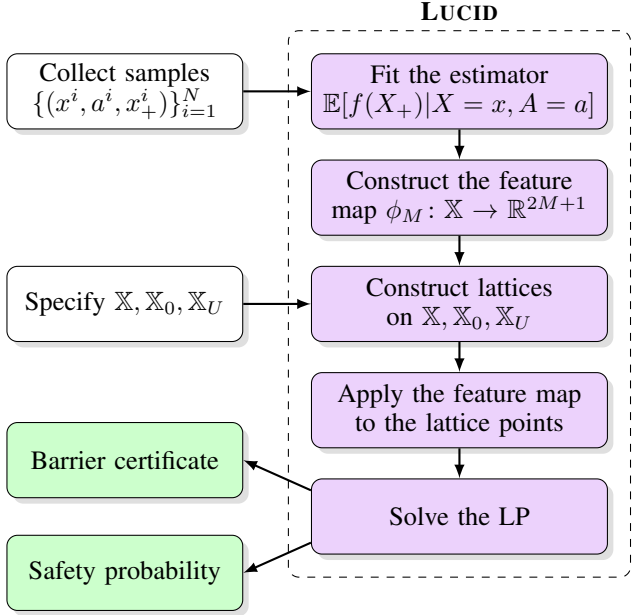
\begin{figure}[ht]
    \centering
    \begin{tikzpicture}[node distance=0.4cm and 0.9cm,
            input-block/.style={rectangle, draw, fill=white, text width=2.9cm, minimum height=1.0cm, align=center, rounded corners, drop shadow=gray!50},
            output-block/.style={rectangle, draw, fill=green!20, text width=2.9cm, minimum height=1.0cm, align=center, rounded corners, drop shadow=gray!50},
            block/.style={rectangle, draw, fill=accent!20, text width=3.7cm, minimum height=1.0cm, align=center, rounded corners, drop shadow=gray!50},
            arrow/.style={thick,->,>=Stealth},
            tool/.style={rectangle, draw, fill=green!20, minimum height=0.8cm, align=center, rounded corners, drop shadow=gray!50},
            inv/.style={},
            arrow/.style={thick,->,>=latex}
        ]

        \node[input-block] (samples) {Collect samples $\{(x^i,a^i, x_+^i)\}_{i=1}^N$};

        \node[block, right=of samples] (estimator) {Fit the estimator
            $\E[f(X_+)\!\mid\! X=x,A=a]$
        };
        \node[block, below=of estimator] (feature) {Construct the feature map $\phi_M\colon \X \to \mathbb{R}^{2M+1}$};
        \node[block, below=of feature] (lattice) {Construct lattices on $\X, \X_0, \X_U$};
        \node[block, below=of lattice] (apply-lattice) {Apply the feature map to the lattice points};
        \node[block, below=of apply-lattice] (optimizer) {Solve the LP};
        \node[draw, dashed, inner sep=0.3cm, rounded corners,
        fit=(estimator)(feature)(lattice)(apply-lattice)(optimizer), label=above:{\textbf{\lucid}}] {};

        \node[input-block, left=of lattice] (specification) {Specify $\X, \X_0, \X_U$};

        \node[output-block, left=of optimizer, yshift=.75cm] (output) {Barrier certificate};
        \node[output-block, left=of optimizer, yshift=-.75cm] (output2) {Safety probability};

        \draw[arrow] (samples) -- (estimator);
        \draw[arrow] (specification) -- (lattice);
        \draw[arrow] (estimator) -- (feature);
        \draw[arrow] (feature) -- (lattice);
        \draw[arrow] (lattice) -- (apply-lattice);
        \draw[arrow] (apply-lattice) -- (optimizer);
        \draw[arrow] (optimizer.170) -- (output.east);
        \draw[arrow] (optimizer.190) -- (output2.east);

    \end{tikzpicture}
    \caption{Sequence of steps \lucid goes through to generate a barrier certificate.}
    \label{fig:steps}
\end{figure}

\paragraph{Safety Problem}
A question that often arises when designing a policy $\pi$ for a system $\S$, especially in engineering contexts, is if the closed-loop system $\S^\pi$ elicits safe behavior. That is, when starting from a domain $\X_{0}\subset\X$ under a policy $\pi$, the system $\S^\pi$ shall avoid unsafe regions $\X_U\subset\X$ (such as obstacles) for at least some predefined time horizon $T\in\N\cup\{\infty\}$.

Certifying safety of a discrete-time stochastic system over an uncountable state space does not generally admit an analytical solution and is extremely challenging, especially for complex dynamics~\cite{abate2008probabilistic}.
Furthermore, for unbounded stochasticity $w_t\sim p_w$, there is generally no yes/no answer to safety, but a safety probability $P_{\text{safe}}(\S^\pi)\in[0,1)$.
Note that given a policy $\pi$, for every initial state $x_0\in\X_0$ there is an associated safety probability. Our goal is to estimate a \emph{lower bound} on the infimum of such probabilities across all initial states in $\X_0$.

\paragraph{Problem Statement}
Assuming access to a finite dataset $\data_N$ from the black-box system $\S$ in \eqref{eq:model}, quantify a certifiable lower bound on the probability of the black-box system $\S^\pi$ being safe with respect to a safety specification given by $(\X_0,\X_U,T)$.

\smallskip

Available sampling-based methods such as Monte Carlo simulation are unfit to solve this problem, as they compute guarantees for fixed initial conditions $x_0\in\X_0$.
\lucid provides a solution for continuous sets $\X_0$ by automating the steps in Figure~\ref{fig:steps}, outlined in the next sections.

\subsection{Control Barrier Certificates}\label{sec:cbc}
\glspl{cbc}\footnote[3]{\glspl{cbc} and discrete-time \glspl{cbf} \cite{cosner2023generative} are equivalent.} leverage the concept of set invariance to arrive at an abstraction-free numerical solution to finding a lower bound on $P_{\text{safe}}(\S^\pi)$. This has made them popular tools for safety verification and controller synthesis~\cite{prajna2006barrier}.

A non-negative function $\B\colon\X\rightarrow\R_{\geq 0}$ is a \gls{cbc} of a system $\S$ with reference to an unsafe set $\X_U$ if it satisfies
\begin{itemize}
    \item[(a)] $\forall x_0\!\in\! \X_0\colon\,\B(x_0)\leq\eta$;
    \item[(b)] $\forall x_U\!\in\! \X_U\colon\,\B(x_U)\geq1$; and
    \item[(c)] $\forall x\!\in\!\X,\,\exists a\!\in\!\A\colon\,\E[\B(X_+)\! \mid\! X=x,\,A=a] -\B(x) \!\leq\!  c;$
\end{itemize}
for some constants $1\!>\!\eta\!\geq\!0$ and $c\!\geq \!0$. Here, upper case $X_+$, $X$, and $A$ denote the random variables underlying concrete realizations $(x_t,a_t,x_{t+1})$ elicited by $\S$.
Intuitively, if one can find a \gls{cbc} for a system $\S$, then, a lower bound on the probability of $\S$ being safe can be quantified based on the distance between the two level sets $1$ and $\eta$~\cite{kushner1967stochastic}:
\begin{equation}
    P_{\text{safe}}(\S^\pi)\geq 1- (\eta + cT),\label{eq:safety_prob_lb}
\end{equation}
where $T$ is the desired time horizon.

Whilst \eqref{eq:safety_prob_lb} can provide a robust assessment of a system's safety, in practice, the bound can be overly conservative and thus several improved variants of the original barrier constraints exist~\cite[e.g.,][]{mahathi2022kinductive}.
As these conditions in essence all rely on the computation of a stochastic constraint with respect to the expected behavior of the system, they are basically interchangeable.

Although there exist model-\emph{based} efficient solutions to this problem for linear and control affine systems, this is generally a semi-infinite problem, demanding a data-driven solution. Furthermore, for the data-driven case, establishing constraint (c) rigorously without relying on impractical assumptions is extremely challenging.

\subsection{Data-Driven Dynamics Estimation via Conditional Mean Embeddings}\label{sec:dd_estimation_via_cme}
To reason about the expected value of a random variable, embedding the variable into a (higher dimensional) space and forming a data-driven estimate is a well-established concept in machine learning \cite{scholkopf2002learning,Steinwart2008SVM}.
Following the same reasoning, the \emph{kernel mean embedding} represents the embedding of a probability measure into an \gls{rkhs} via a feature map $\phi$ associated with a positive definite kernel $k$~\cite{Smola2007EmbedDistrb,muandet2017kernel}.
For conditional probability distributions a similar concept exists: \glspl{cme} can be used to model the expected value of any \gls{rkhs} function $g\colon\X\rightarrow\R$ under a stochastic process such as $\S$~\cite{park2020measuretheoretic,muandet2017kernel}.
For the Gaussian kernel,
\begin{equation}
    k(x,x') := \sigma_f^2 \exp\left( -\tfrac{1}{2} (x-x')\T \Sigma\, (x-x') \right),\label{eq:gaussian_kernel}
\end{equation}
where $\Sigma:=\diag(\sigma_l)^{-2}$, with hyperparameters $\sigma_f,\sigma_l\in\R$, the associated RKHS encompasses all smooth functions $g$.
A data-driven estimate can then be obtained in closed form from a finite amount of data $\data_N$:
\begin{align}
    \begin{split}
        & \E[f(X_+)\mid X=x,\,A=a] \approx                                                  \\
        & \hspace{5em} k_{XA}^N(x,a)\T\left[ K_{XA}^N+N \lambda I_N\right]^{-1}\! f(X_+^N),
    \end{split}\label{eq:empiricalEstimate}
\end{align}
with column vector $k_{XA}^N(x,a):=[k((x^i,a^i),(x,a))]_{i=1}^N$, Gram matrix $K_{XA}^N:=[k((x^i,a^i),(x^j,a^j))]_{i,j=1}^N$, regularization factor $\lambda\geq0$, identity matrix $I_N$, and $f(X_+^N):=[f(x^i_+)]_{i=1}^N$.
The empirical estimator in \eqref{eq:empiricalEstimate} converges in expectation to the true \gls{cme} for $N\rightarrow\infty$ and $\lambda\to0$~\cite{park2020measuretheoretic}.

It is common practice to robustify empirical estimates such as \eqref{eq:empiricalEstimate} to out-of-sample behavior by constructing an \gls{rkhs} ambiguity set centered at the empirical \gls{cme}. The result is a distributionally robust estimator with an adjustable robustness radius. The details are omitted here for brevity, but the interested reader is referred to the works by \citet{kuhn2025distributionally} and \citet{li2022optimal}.

\subsection{Data-Driven Spectral Barriers}\label{sec:ddbarriers}
\begin{figure}
    \centering
    \includegraphics[width=0.9\linewidth]{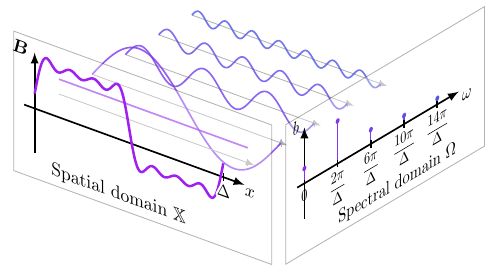}
    \caption{Spectral barrier certificate $\B(x)=\phi_M(x)\T b$.}
    \label{fig:spectral_barrier}
\end{figure}

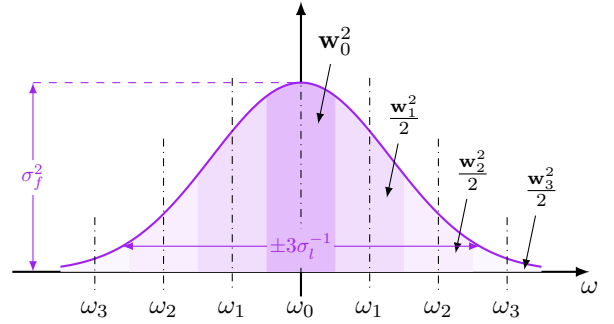
\begin{figure}
    \centering
    \begin{tikzpicture}
        \def\N{50} %
        \def\B{0};
        \def\Bs{3.0};
        \def\var{1.3}
        \def\xmax{\B+3.5*\Bs};
        \def\ymin{{-0.1*gauss(\B,\B,\Bs)}};
        \def\h{0.08*gauss(\B,\B,\Bs)};

        \begin{axis}[every axis plot post/.append style={
            mark=none,domain={-(\xmax)}:{1.0*\xmax},samples=\N,smooth},
            xmin={-(\xmax)}, xmax=\xmax,
            axis/.style={>=latex},
            ymin=\ymin, ymax={1.1*gauss(\B,\B,\Bs)},
            axis lines=middle,
            axis line style=thick,
            enlarge x limits, %
            ticks=none,
            xlabel=$\omega$,
            every axis x label/.style={at={(current axis.right of origin)},anchor=north},
            width=1.1\linewidth, height=0.55*\linewidth,
            y=700pt,
            clip=false,
            axis line style={-latex}
            ]

            \addplot[accent,thick,name path=B] {gauss(x,\B,\var*\Bs)};

            \path[name path=xaxis]
            (\B-\pgfkeysvalueof{/pgfplots/xmax},0) -- (\B+\pgfkeysvalueof{/pgfplots/xmax},0); %
            \addplot[accent!3.75] fill between[of=xaxis and B, soft clip={domain={\B-3.5*\Bs}:{\B+3.5*\Bs}}];
            \addplot[accent!7.5] fill between[of=xaxis and B, soft clip={domain={\B-2.5*\Bs}:{\B+2.5*\Bs}}];
            \addplot[accent!15] fill between[of=xaxis and B, soft clip={domain={\B-1.5*\Bs}:{\B+1.5*\Bs}}];
            \addplot[accent!30] fill between[of=xaxis and B, soft clip={domain={\B-.5*\Bs}:{\B+.5*\Bs}}];

            \addplot[black,dashdotted,thin]
            coordinates {({\B-3*\Bs},{20*gauss(\B-3*\Bs,\B,\Bs)}) ({\B-3*\Bs},{-\h})}
            node[below=-3pt,scale=1.0] {\strut$\omega_3$};
            \addplot[black,dashdotted,thin]
            coordinates {({\B-2*\Bs},{4*gauss(\B-2*\Bs,\B,\Bs)}) ({\B-2*\Bs},{-\h})}
            node[below=-3pt,scale=1.0] {\strut$\omega_2$};
            \addplot[black,dashdotted,thin]
            coordinates {({\B-1*\Bs},{1.3*gauss(\B-\Bs,\B,\Bs)}) ({\B-1*\Bs},{-\h})}
            node[below=-3pt,scale=1.0] at ({\B-\Bs},{-\h}) {\strut$\omega_1$};
            \addplot[black,dashdotted,thin]
            coordinates {(\B,{1.05*gauss(\B,\B,\Bs)}) (\B,{-\h})}
            node[below=-3pt,scale=1.0] {\strut$\omega_0$};
            \node[scale=1.0] (w0)
            at ({\B+.5*\Bs},{.125}) {$\w_0^2$};
            \node (b0) at ({\B+.25*\Bs},{.08}) {};
            \draw[->] (w0) to (b0.center);
            \addplot[black,dashdotted,thin]
            coordinates {({\B+1*\Bs},{1.3*gauss(\B+\Bs,\B,\Bs)}) ({\B+1*\Bs},{-\h})}
            node[below=-3pt,scale=1.0] at ({\B+\Bs},{-\h}) {\strut$\omega_1$};
            \node[scale=1.0] (w1)
            at ({\B+1.5*\Bs},{.085}) {$\frac{\w_1^2}{2}$};
            \node (b1) at ({\B+1.25*\Bs},{.04}) {};
            \draw[->] (w1) to (b1.center);
            \addplot[black,dashdotted,thin]
            coordinates {({\B+2*\Bs},{4*gauss(\B+2*\Bs,\B,\Bs)}) ({\B+2*\Bs},{-\h})}
            node[below=-3pt,scale=1.0] at ({\B+2*\Bs},{-\h}) {\strut$\omega_2$};
            \node[scale=1.0] (w2)
            at ({\B+2.5*\Bs},{.055}) {$\frac{\w_2^2}{2}$};
            \node (b2) at ({\B+2.25*\Bs},{.01}) {};
            \draw[->] (w2) to (b2.center);
            \addplot[black,dashdotted,thin]
            coordinates {({\B+3*\Bs},{20*gauss(\B+3*\Bs,\B,\Bs)}) ({\B+3*\Bs},{-\h})}
            node[below=-3pt,scale=1.0] at ({\B+3*\Bs},{-\h}) {\strut$\omega_3$};
            \node[scale=1.0] (w3)
            at ({\B+3.5*\Bs},{.046}) {$\frac{\w_3^2}{2}$};
            \node (b3) at ({\B+3.25*\Bs},{.001}) {};
            \draw[->] (w3) to (b3.center);

            \addplot[<->,accent]
            coordinates {({\B-2*\var*\Bs},{gauss(\B+2*\var*\Bs,\B,\var*\Bs)}) ({\B+2*\var*\Bs},{gauss(\B+2*\var*\Bs,\B,\var*\Bs)})};
            \node[accent,fill=accent!30,inner xsep=1,inner ysep=2,scale=.8] at (\B,{gauss(\B+2*\var*\Bs,\B,\var*\Bs)}) {$\pm 3\sigma_l^{-1}$};
            \addplot[<->,accent]
            coordinates {({\B-3*\var*\Bs},{gauss(0,0,\var*\Bs)}) ({\B-3*\var*\Bs},{0})};
            \addplot[dashed,thin,accent]
            coordinates {({\B-3*\var*\Bs},{gauss(0,0,\var*\Bs)}) ({0},{gauss(0,0,\var*\Bs)})};
            \node[accent,fill=white,inner xsep=3,inner ysep=2,scale=.8] at (\B-3*\var*\Bs,{.5*gauss(0,0,\var*\Bs)}) {$\sigma_f^2$};

        \end{axis}
    \end{tikzpicture}
    \caption{Abstraction of a 1-dim. Gaussian spectral measure of the Gaussian kernel, shown in \eqref{eq:sqexp_kernel_fourier}.}
    \label{fig:spectral_measure_abstraction}
\end{figure}

Based on the data-driven estimator in \eqref{eq:empiricalEstimate}, the problem of computing \glspl{cbc} using data can be formulated as a nonconvex semi-infinite program, which for general classes of systems and barriers is extremely difficult to solve.
To arrive at a tractable solution, \lucid conducts two additional steps:
\paragraph{1. Spectral Abstraction} Inspired by the popular random Fourier features approach by \citet{rahimi2007random}, the Gaussian kernel \eqref{eq:gaussian_kernel} admits a Fourier expansion
\begin{align}
    k(x, x') \equiv \sigma_f^2 \int_{\R^{n}}\!\! \mathcal{N}(d\omega\mid0,\Sigma)\; e^{\mathbf{i}\omega\T (P(x)-P(x'))},\label{eq:sqexp_kernel_fourier}
\end{align}
with the zero-mean Gaussian distribution $\smash{\mathcal{N}(d\omega\mid0,\Sigma)}$ with covariance $\Sigma$, the imaginary unit $\mathbf{i}:=\smash{\sqrt{-1}}$, and where the affine transform $x \mapsto P(x)$ maps the domain $\X$ into the unit hypercube $[0,1]^{n}$.
Partitioning the space of spatial frequencies $\omega\in\R^n$ into a set of discrete frequency bands (see Figure~\ref{fig:spectral_barrier}) yields a spectral abstraction of the associated \gls{rkhs}, i.e., characterizing learnable functions $\B$ in the form of Fourier series.
By truncating the series to a finite number, $M$, of fixed frequency bands $\omega_j\in\R^{n}$, $j\in\{0,\ldots,M\}$, the resulting barriers are of the form
\begin{align*}
    \B(x) = \alpha_0 +\sum_{i=1}^{M} \alpha_i \cos\left(\omega_i\T P(x)\right) + \beta_i \sin\left(\omega_i\T P(x)\right).%
\end{align*}
Notably, $\B$ admits the linear form $\B(x)=\phi_M(x)\T b$, with a truncated Fourier feature map $\phi_M\colon\X\rightarrow\R^{2M+1}$, parametrized by learnable spectral amplitudes
\begin{align*}
    b\!=\!\begin{bmatrix}
              \frac{\alpha_0}{\sigma_f^2\w_0^2}\!\! & \frac{\alpha_1}{2\sigma_f^2\w_1^2}\!\!\! & \frac{\beta_1}{2\sigma_f^2\w_1^2}\!\!\! & \ldots\!\!\! & \frac{\alpha_{M}}{2\sigma_f^2\w_{M}^2}\!\!\! & \frac{\beta_{M}}{2\sigma_f^2\w_{M}^2}
          \end{bmatrix}\T\!\!\!\!\in\R^{2M+1}\!\!,
\end{align*}
where the weights $\w_0,\ldots,\w_{M}\in\R_{\geq 0}$ associated with each frequency band are determined efficiently from the kernel's Gaussian spectral measure via the multivariate CDF (see Figure~\ref{fig:spectral_measure_abstraction}).
The \gls{cme}-based estimator \eqref{eq:empiricalEstimate} is approximated in the same finite basis via $H\in\R^{(2M+1)\times (2M+1)}$ such that
\begin{equation*}
    k_{XA}^N(x,a)\T\left[ K_{XA}^N+N \lambda I_N\right]^{-1}\! \Phi_{M,+}^N
    \approx
    \varphi_M(x,a)\T H,
\end{equation*}
where $\Phi_{M,+}^N:=[\phi_M(x_+^i)\T]_{i=1}^N$, $\varphi_M(x,a)\colon\X\times\A\rightarrow\smash{\R^{2M+1}}$ a feature map augmenting $\phi_M(x)$, and
with an approximation error decreasing exponentially with $M$~\cite{rahimi2007random}.
This reduces the barrier synthesis problem
to a semi-infinite linear program, with barrier conditions linear in $b$:
\begin{itemize}
    \item[(a)] $\forall x_0\in \X_0\colon\,\phi_M(x_0)\T b\leq\eta$;
    \item[(b)] $\forall x_U\in \X_U\colon\,\phi_M(x_U)\T b\geq1$; and
    \item[(c)] $\forall x\in\X,\,\exists a\in\A\colon\,\varphi_M(x,a)\T (Hb -b) \leq  c.$
\end{itemize}
For the case where $a$ is provided by a given fixed policy $\pi$, $\varphi_M(x,a)$ reduces to $\phi_M(x)$.

\paragraph{2. Finite-Constraint Relaxation} To obtain a program with \emph{finitely} many constraints, trigonometric bounding results \cite{pfister2018bounding,schoenJAIR} are used to obtain a relaxed linear program by sampling spatial lattices on $\X$, $\X_0$, and $\X_U$.
For example, to enforce barrier constraint (a) in Section~\ref{sec:cbc}, it suffices to construct a lattice $\{x_0^i\}_{i=1}^{N_0}\subset\X_0$ and impose the constraints $b\T\phi_M(x_0^i)\leq\eta-\epsilon$ for $i=1,\ldots,N_0$, where $\epsilon>0$ is a computed coefficient.
Selecting lattices with a sampling density meeting the Nyquist-Shannon sampling theorem with respect to the highest frequency $\omega_M$ appearing in the barrier and truncated estimator, the relaxation retains all guarantees.
As a result, the semi-infinite problem is relaxed to a finitely-constrained \gls{lp}, provided in full in the
\if\extendedversion1
    appendix.
\else
    extended version~\cite{extended_version}.
\fi
Recall that given an appropriate robustness radius, the barriers produced by \lucid based on data $\data_N$ are statistically correct with respect to the unknown true system $\S$, and a lower bound for the safety probability is computed according to \eqref{eq:safety_prob_lb}.

\section{Tool Structure and Functionalities}
\label{sec:tool}
\lucid implements the previously outlined functionality, written in C++ and designed to be used as a library or standalone executable.
The choice of a low level language gives us plenty of freedom and fine-grained control over the execution of the software.
We expose a set of interfaces allowing users to highly customize the verification process.
A high-level visualization of \lucid's architecture is illustrated in Figure~\ref{fig:architecture}, while a more technical description can be found in the online documentation at
\begin{center}
    \url{https://tendto.github.io/lucid/}.
\end{center}

We also provide a Python wrapper, called \pylucid, to facilitate the integration of the tool into existing workflows and effortlessly leverage well-established libraries such as NumPy~\cite{numpy} and SciPy~\cite{scipy}.
Moreover, \pylucid can be extensively configured in a variety of ways (e.g., Python scripts, YAML files, GUI), making it the recommended way to operate \lucid.
For the rest of the paper, we will thus focus on \pylucid when describing the user interfaces.
Further information about \pylucid are deferred to the
\if\extendedversion1
    appendix.
\else
    extended version~\cite{extended_version}.
\fi

\paragraph{Configuration}
To certify safety of a system, \lucid accepts a configuration comprising data from the system and the safety specification (see Figure~\ref{fig:architecture}).
We suggest defining it as a \yaml or equivalent \json file.
If more flexibility is needed, a Python script generating a configuration can be used instead.
Regardless of their format, configuration files can be loaded with the command \lstinline|pylucid <config file>| to start the tool's verification pipeline.
The same configuration can also be passed directly as command line arguments.
\pylucid also provides a browser-based \gls{gui} to aid in the configuration and execution of the tool.

In its current form, \lucid implements all the functionality needed to certify safety of a closed-loop system with a given control policy $\pi$.
Thus the action $a$ in the previous section is replaced with the policy $\pi(x)$.
Future releases will expand this core functionality to
safe controller synthesis as well.
Due to its modular architecture, \lucid can be easily extended in multiple directions, as outlined in Section~\ref{sec:conclusion_and_future_extensions}.

\subsection{Configuring and Running the Tool}\label{sec:config_and_run_tool}

\begin{figure}
    \centering
    \begin{tikzpicture}[node distance=0.6cm and 0.6cm,
            input-block/.style={rectangle, draw, fill=red!10!white, text width=2.5cm, minimum height=0.8cm, align=center, rounded corners, drop shadow=gray!50},
            output-block/.style={rectangle, draw, fill=green!20!white, text width=2.6cm, minimum height=0.8cm, align=center, rounded corners, drop shadow=gray!50},
            block/.style={rectangle, draw, fill=accent!20!white, text width=2.0cm, minimum height=0.8cm, align=center, rounded corners, drop shadow=gray!50},
            arrow/.style={thick,->,>=latex},
            doc/.style={draw,thick,align=center,color=black,shape=document,minimum width=20mm-6mm,minimum height=24.2mm-10mm,shape=document,inner sep=1ex,fill=white,drop shadow=gray!50},
            inv/.style={}
        ]

        \node[doc] (input) {Configuration\\[.2em]\textit{.yaml/.json/.py}};

        \node[block, below=1cm of input] (estimator) {\Circled{2} Estimator};
        \node[block, right=of estimator] (tuner) {\Circled{3} Tuner};
        \node[block, left=of estimator] (sets) {\Circled{1} Safety Specification};
        \node[block, below=of estimator, text width=2.4cm] (feature) {\Circled{4} Feature Map}; %
        \node[block, left=.4cm of feature] (optimizer) {\Circled{5} Solver};
        \node[draw, dashed, inner sep=0.25cm, rounded corners,
        fit=(estimator)(feature)(optimizer)(sets)(tuner), label={[xshift=3cm]below:{\textbf{\lucid}}}] {};

        \node[output-block, below=of optimizer] (output) {Barrier Certificate};
        \node[output-block, right=.5cm of output] (prob) {Safety Probability};
        \node[] at ($(output)!0.5!(prob)$) (plus) {$+$};

        \draw[arrow] (input) -- (estimator);
        \draw[arrow] ($(input.south)+(0.0,-0.3)$) -| (tuner);
        \draw[arrow] ($(input.south)+(0.0,-0.3)$) -| (sets);
        \draw[arrow] (sets) -- (optimizer);
        \draw[arrow] (tuner) -- (estimator);
        \draw[arrow] (estimator) -- (feature);
        \draw[arrow] (feature) -- (optimizer);
        \draw[arrow] (optimizer) -- (output);

    \end{tikzpicture}
    \caption{General architecture of \lucid, highlighting its core components and their connections.}
    \label{fig:architecture}
\end{figure}
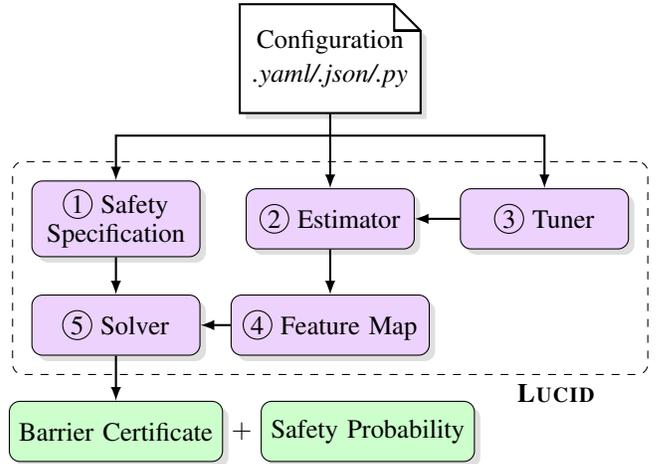

\lucid is built with a modular architecture, where each component is responsible for a specific task in the certification process.
Since the components extend from a common interface, they can be easily replaced or extended.
An overview of the core components and how they interact is shown in Figure~\ref{fig:architecture}.
Classes and interfaces available in \pylucid are written in \texttt{monospace \!\!font}.

To make the explanation more intuitive, we will use a one-dimensional linear system as a running example:
\begin{equation}
    \label{eq:linear-system}
    \S\colon%
    \quad
    {x}_{t+1}
    =
    0.5
        {x}_{t}
    + w_t,
    \quad w_t \sim \mathcal{N}(\cdotx | 0, 0.01),
\end{equation}
with
state space $\X = [-1, 1]$, initial set $\X_0 = [-0.5, 0.5]$, and unsafe regions $\X_U = [-1, -0.9] \cup [0.9, 1]$.
While going over each component of \lucid, we will show how they can be configured within a \yaml configuration file to capture the system~\eqref{eq:linear-system} and its safety specification.

\paragraph{Dataset}
Due to its data-driven nature, all that \lucid needs to operate is a set of sampled transitions $\data_N$ from the system, as shown in \eqref{eq:data}, and the safety specification itself.
The samples can be specified directly in the configuration file, divided into $x^1,\ldots,x^N$ and $x^1_+,\ldots,x^N_+$:
\begin{lstlisting}[language=yaml,backgroundcolor=\color{ipython_bg}]
x_samples: [[5.930e-01], ..., [-1.937e-01]]
xp_samples: [[3.015e-01], ..., [-1.018e-01]]
\end{lstlisting}
Alternatively, the samples can be either generated by the Python configuration script or stored in a separate file.

\paragraph{\Circled{1}~Safety Specification}
\lucid understands sets $\X,\X_0,\X_U$ specified as \texttt{RectSet}, \texttt{SphereSet}, or \texttt{Multiset} objects (collections of sets from the first two categories).
They can be included in the configuration as shown below:
\begin{lstlisting}[language=yaml,backgroundcolor=\color{ipython_bg}]
X_bounds: "RectSet([-1], [1])"
X_init: "RectSet([-0.5], [0.5])"
X_unsafe: ["RectSet([-1], [-0.9])",
           "RectSet([0.9], [1])"]
\end{lstlisting}

\paragraph{\mbox{\Circled{2}\hspace{.4em}Estimator}}
The core of \lucid is its \estimator, namely the \texttt{KernelRidgeRegressor}, which uses the \texttt{GaussianKernel} to learn the underlying system dynamics by estimating the \gls{cme} from the samples.
Its predictions of the expected next state $x_+$ are used
to determine the constraints for the \gls{cbc} \gls{lp}.
The setup is configured as follows:
\begin{lstlisting}[language=yaml,backgroundcolor=\color{ipython_bg}]
kernel: "GaussianKernel"
estimator: "KernelRidgeRegressor"
\end{lstlisting}

\paragraph{\Circled{3}~Tuner}
Being parameter-free approaches, kernel methods do not require the expensive learning processes of other machine learning methods, such as neural networks.
However, they still depend on a number of \hp, such as the kernel bandwidth $\sigma_f$, the lengthscale $\sigma_l$, and the regularization constant $\lambda$ (see \eqref{eq:gaussian_kernel}--\eqref{eq:empiricalEstimate}).
Changes in their values can have significant impact on the \estimator's efficiency and accuracy.
The process of finding good values for these \hp is known as \emph{hyperparameter tuning}, with the optimal parameters being problem dependent.
\lucid provides a set of utilities, which specialize the \texttt{Tuner} interface, to aid in this task:
\begin{itemize}
    \item \texttt{MedianHeuristicTuner} uses the \emph{median heuristic} \cite{garreau2018largesampleanalysismedian} to produce rule-of-thumb-type estimates for the \hp $\sigma_f$ and $\sigma_l$ of, e.g., the Gaussian kernel \eqref{eq:gaussian_kernel}, via closed-form expressions.
          It can be useful as a starting point for subsequent improvements.
    \item \texttt{LbfgsTuner} finds the \hp that maximize the \emph{log marginal likelihood}~\cite{rasmussen2006gaussian}, defined as
          \begin{equation*}
              \begin{split}
                  & \log p(X^+_N | X_N,\theta) =  - \frac{1}{2}X^+_N{}\T (K_{X}^N + N\lambda I_N)^{-1} X^+_N \\
                  & \hspace{40pt} -\frac{1}{2}\log |K_{X}^N + N\lambda I_N| - \frac{N}{2}\log(2\pi),
              \end{split}
          \end{equation*}
          where $\theta:=(\sigma_f,\sigma_l,\lambda)$ is the hyperparameterization being optimized.
          We use the L-BFGS or L-BFGS-B quasi-Newton optimization algorithms~\cite{book:nonlinear-programming},
          implemented in the \texttt{LBFGS++} library.
    \item \texttt{GridSearchTuner} implements the grid search method, exploring the space of possible hyperparameter values to maximize the \estimator's $R^2$ score,  $$R^2 = 1 - \textstyle\left(\sum_{i=1}^M (y_i - \hat{y}_i)^2/\sum_{i=1}^M (y_i - \bar{y})^2\right),$$ with $\hat{y}_i$ and $\bar{y}:=(\sum_{i=1}^N y_i)/N$ being the $i^{\text{th}}$ predicted and mean observed outputs, respectively.
\end{itemize}
Tuners open a wide range of possibilities to the user.
We recommend using them within a Python script generating a configuration to automate the tuning process before finalizing the \estimator's hyperparameter values:
\begin{lstlisting}[language=iPython,style=nonumbers]
def scenario_config(c: Configuration):
    t = LbfgsTuner(lb=[1e-5], ub=[1e5])
    c.estimator = KernelRidgeRegressor()
    c.estimator.fit(c.x_samples, c.xp_samples, tuner=t)
    return c
\end{lstlisting}
In this example, we set the \hp explicitly:
\begin{lstlisting}[language=yaml,backgroundcolor=\color{ipython_bg}]
sigma_l: 0.0446
lambda: 1.0e-5
set_scaling: 0.04
\end{lstlisting}
The parameter \lstinline{set_scaling} can be used to lower the constraint-tightening $\epsilon$ (see Section~\ref{sec:ddbarriers}.2.) by increasing the size of the sets $\X,\X_0,\X_U$; here, for example, by 4\%.

\paragraph{\Circled{4}~Feature Map}
We exploit the spectral kernel expansion in \eqref{eq:sqexp_kernel_fourier} to construct an explicit approximated feature map, composed of trigonometric functions with increasing frequencies (see Figure~\ref{fig:spectral_barrier}).
The underlying kernel expansion, visualized in Figure~\ref{fig:spectral_measure_abstraction}, can be controlled to trade-off efficiency and accuracy/conservativeness.
After selecting its \hp $(\sigma_f,\sigma_l)$, the feature map can be used to map any point from $\X$ to the \gls{rkhs} associated with the kernel.
To this end, \lucid partitions the kernel's spectral measure:
\texttt{LinearTruncatedFourierFeatureMap} defines equally spaced frequency bands on the interval $[-3\sigma_l,3\sigma_l]$, capturing 99.73\% of the spectral measure. This partitioning is shown in Figure~\ref{fig:spectral_measure_abstraction}.
Users may define custom feature maps to produce other partitions.

In our example, we truncate the Fourier expansion to $6$ frequency bands, including the constant term, and use $300$ lattice points to build the constraints for the optimization.
\begin{lstlisting}[language=yaml,backgroundcolor=\color{ipython_bg}]
num_frequencies: 6
lattice_resolution: 300
feature_sigma_l: 0.0925
\end{lstlisting}

\paragraph{\Circled{5} Solver}
From the components discussed above, \lucid generates and solves a finite-constraint \gls{lp} as outlined in Section~\ref{sec:ddbarriers}.
If successful, \lucid returns a \gls{cbc}, a quantified lower bound on the true safety probability $P_{\text{safe}}(\S^\pi)$, as well as the corresponding constants $\eta$ and $c$.

Generating the \gls{lp}, specifically choosing an appropriate lattice density, involves trading off efficiency versus conservativeness, with any sampling density beyond the Nyquist frequency yielding a valid relaxation.
In the benchmarks in Section~\ref{sec:experiments} we showcase the relation numerically.
For solving the \gls{lp}, \lucid can interface with different linear optimizers: \gurobi, \alglib, or \highs. Here, we use \gurobi:
\begin{lstlisting}[language=yaml,backgroundcolor=\color{ipython_bg}]
optimiser: "GurobiOptimiser"
\end{lstlisting}

\paragraph{Running the Tool}
We have configured all the components needed to run \lucid.
Putting it all together in a single configuration file, we can run the tool with the command \lstinline|pylucid config.yaml --plot|.
\lucid will parse the configuration, synthesize a barrier certificate, and plot the result.
For the running example, we obtain the barrier certificate shown in Figure~\ref{fig:example} for a time horizon of $T = 15$, certifying safety of the system in \eqref{eq:linear-system} with a probability of at least $93.07\%$ ($\eta = 0.006, c = 0.004$).

\begin{figure}
    \centering
    \includegraphics[width=1\linewidth]{linear.pdf}
    \caption{Barrier certificate for the running example.
        The value of the barrier $\B$ is plotted against the state space $\X$.
        The green line represents the barrier value.
        The initial and unsafe sets are highlighted in blue and red, respectively. %
    }
    \label{fig:example}
\end{figure}

\paragraph{Validation (optional)}
Albeit barriers generated by \lucid are by design formally sound with respect to the data-driven estimator,
if the latent system dynamics are known, \lucid provides the option to formally verify the correctness of the resulting barrier using the \dreal SMT solver~\cite{dreal}.
Since we know the expected behavior of the system $\S$ from~\eqref{eq:linear-system}, we confirm that $\B$ is indeed a valid \gls{cbc} by adding the following lines:
\begin{lstlisting}[language=yaml,backgroundcolor=\color{ipython_bg}]
system_dynamics: ["x1 / 2"]
verify: true
\end{lstlisting}

\section{Experimental Evaluation}
\label{sec:experiments}

To ascertain the performance of \lucid, we conduct a series of experiments on a Windows 10 machine with an AMD Ryzen 9 5950X 16-Core Processor @ 3.40 GHz, NVIDIA GeForce RTX 3090 GPU, and 64 GB of RAM.
All runs had the random seed set to $42$ to ensure reproducibility.

\begin{figure}[ht]
    \centering
    \includegraphics[width=1\linewidth]{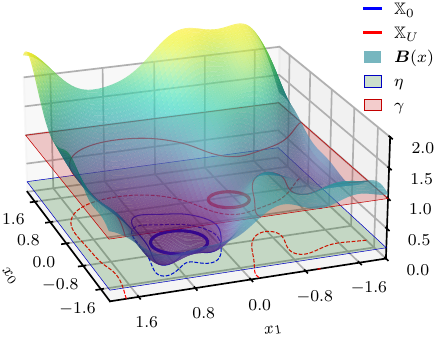}
    \caption{\gls{cbc} synthesized for the \barrII benchmark.}
    \label{fig:CSBarr2}
\end{figure}

\begin{figure}[ht]
    \centering
    \includegraphics[width=1\linewidth]{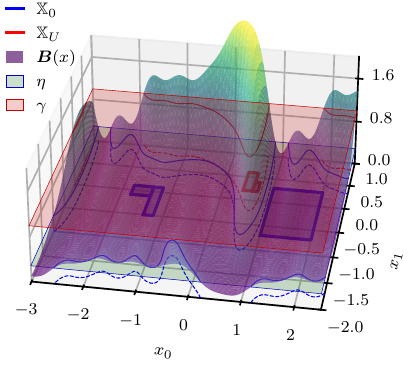}
    \caption{\gls{cbc} synthesized for the \barrIII benchmark.}
    \label{fig:CSBarr3}
\end{figure}

We adapt the \barrII and \barrIII benchmarks from \citet{abate2021fossil}.
Both are two-dimensional highly nonlinear systems to which we add stochastic noise $w_t\sim\mathcal{N}(\cdotx\vert 0,0.01I_2)$.
Given $N=1000$ samples, initial set $\X_0$, and unsafe set $\X_U$, we synthesize a barrier $\B$ that guarantees trajectories starting in $\X_0$ do not enter $\X_U$ within $T=5$ time steps.
Note that here we only certify safety w.r.t. the empirical distribution, i.e., the \gls{cme} constructed from the observed data.
The synthesized barriers are shown in Figures~\ref{fig:CSBarr2}--\ref{fig:CSBarr3}.
We also consider a new benchmark \overtaking, where an autonomous vehicle controlled by a \gls{nn} is overtaking another vehicle.
The dynamics of the ego vehicle are given by Dubin's car model with an added noise vector $w$ where each component is drawn from a zero-mean Gaussian with standard deviation $0.01$, $0.01$, and $0.001$ respectively.
The steering wheel angle is supplied by the \gls{nn} controller and we travel at a fixed velocity.

Table~\ref{tbl:benchmarks_results} summarizes the results of all the experiments presented.
As a point-wise baseline, we estimate the safety probability at selected initial states via Monte Carlo simulation, yielding approximately 95--100\% safety in the reported benchmarks. Note that these estimates do not extend to set-wise guarantees.
Further details on the experiments can be found in the
\if\extendedversion1
    appendix.
\else
    extended version~\cite{extended_version}.
\fi

\begin{table}[tb]
    \centering
    \begin{tabular}{cccccc}
        \toprule
                    & \textbf{T} & \textbf{\#Freq.} & \textbf{Lattice} & \textbf{Runtime} & \textbf{Safety} \\
                    &            &                  & \textbf{Size}    & [mm:ss]          & \textbf{Prob.}  \\ %
        \midrule
        \linear     & 15         & 6                & $300^1$          & 00:01            & 93.07\%         \\
        \barrII     & 5          & 7                & $330^2$          & 04:38            & 63.17\%         \\
        \barrIII    & 5          & 7                & $330^2$          & 03:20            & 51.63\%         \\
        \overtaking & 5          & 5                & $70^3$           & 93:56            & 53.66\%         \\
        \bottomrule
    \end{tabular}
    \caption{Computational benchmarks:
        The lattice size indicates the number of points of the lattice in each dimension.
        $T$ is the time horizon associated with the lower bound on the safety probability, displayed in the last column.
    }
    \label{tbl:benchmarks_results}
\end{table}

\section{Conclusion and Future Extensions}\label{sec:conclusion_and_future_extensions}
This paper introduces \lucid, a novel tool capable of quantifying safety guarantees for black-box systems with complex stochastic dynamics and deep learning components in the loop. As such, \lucid fills a gap currently unaddressed by preexisting tools. To achieve this, \lucid leverages Conditional Mean Embeddings (CME) to learn control barrier certificates from data and recasts the problem into a tractable linear form via a spectral abstraction.

\lucid is built for extensibility. In its current form, it assumes access to full state measurements (see~\eqref{eq:data}), as is the case when working with simulated systems, and focuses on verification, i.e., when a policy is given.
Extending \lucid to partially observed settings and safe controller synthesis
is on our agenda and builds on top of the results presented in this paper.
\lucid derives barriers based on the empirical CME, which can be robustified against out-of-sample system behavior by tightening the barrier constraint (c) (see Sections~\ref{sec:dd_estimation_via_cme}--\ref{sec:ddbarriers}).
The scalability of \lucid can be improved by performing sparse CME computations or specializing the estimator's \texttt{Kernel} to embed prior knowledge about the system dynamics.

\section*{Acknowledgments}
Ernesto Casablanca is supported by the Engineering and Physical Sciences Research Council (EPSRC), grant number EP/W524700/1.
Paolo Zuliani is supported by the project SERICS (PE00000014) under the Italian MUR National Recovery and Resilience Plan funded by the European Union - NextGenerationEU.
The work of Sadegh Soudjani is supported by the EIC SymAware project 101070802 and the ERC Auto-CyPheR project 101089047.

\bibliography{aaai25}

\if\extendedversion1
    \appendix
    \appendix
\onecolumn

\section{Linear Program}

The finitely-constrained LP discussed in Section~\ref{sec:ddbarriers} is presented. For this, the lattices $\smash{\Theta_{\tilde{N}}:=\{ x^{1}, \ldots, x^{\tilde{N}} \}}\subset\tilde{\X}$, $\smash{\{ x^{1}, \ldots, x^{{N}} \}}\subset\X$, $\smash{\{ x_0^{1}, \ldots, x_0^{{N}_0} \}} \subset \X_0$, and $\smash{\{ x_U^{1}, \ldots, x_U^{N_U} \}} \subset \X_U$ of cardinality $\tilde{N},N,{N}_0,N_U\in\N$, respectively, are formed.
For given values of $\bar{\B}$ and robustness radius $\mathcal{\varepsilon}\geq0$, the following LP is obtained:
\begin{equation}
      \begin{alignedat}{3}
            & \min_{\substack{b, c, \eta                                                                                                                                                                                \\
            \Bmin_{\tilde{N}}^{\X_0},\Bmax_{\tilde{N}}^{\X_u},\Bmin_\Delta^{\X},\Bmax_{\tilde{N}}^{{\X}}                                                                                                             \\
            \Bmax_{\tilde{N}}^{\tilde{\X}\setminus\X_0},\Bmin_{\tilde{N}}^{\tilde{\X}\setminus\X_u},\Bmin_{\tilde{N}}^{\tilde{\X}\setminus\X},\Bmax_\Delta^{\tilde{\X}\setminus\X}
            }} \quad &                            & \eta + cT,                                                                &                                                                        &                         \\
            & \text{subject to}\quad
            &                            & \Bmin_{\tilde{N}}^{\X_0}\leq\phi_M(x_0^{(i)})\T b\leq\hat{\eta}, \quad &                                                                        & i=1,\ldots,\hat{N}_0,   \\
            &                            &                                                                           & \hat{\gamma}\leq\phi_M(x_u^{(i)})\T b\leq\Bmax_{\tilde{N}}^{\X_u},
            \quad    &                            & i=1,\ldots,\hat{N}_u,                                                                                                                                                        \\
            &                            &                                                                           & \Bmin_\Delta^{\X}\leq\phi_M(x^{(i)})\T(Hb - b) \leq \hat{\Delta},
            \quad    &                            & i=1,\ldots,\hat{N},                                                                                                                                                          \\
            &                            &                                                                           & \hat{\xi}\leq\phi_M(x^{(i)})\T b\leq\Bmax_{\tilde{N}}^{{\X}},
            \quad    &                            & i=1,\ldots,\hat{N},                                                                                                                                                          \\
            &                            &                                                                           & \phi_M(x^{(i)})\T b\leq\Bmax_{\tilde{N}}^{\tilde{\X}\setminus\X_0},
            \quad    &                            & i=1,\ldots,\tilde{N}-\hat{N}_0,                                                                                                                                              \\
            &                            &                                                                           & \Bmin_{\tilde{N}}^{\tilde{\X}\setminus\X_u}\leq\phi_M(x^{(i)})\T b,
            \quad    &                            & i=1,\ldots,\tilde{N}-\hat{N}_u,                                                                                                                                              \\
            &                            &                                                                           & \Bmin_{\tilde{N}}^{\tilde{\X}\setminus\X}\leq\phi_M(x^{(i)})\T b,
            \quad    &                            & i=1,\ldots,\tilde{N}-\hat{N},                                                                                                                                                \\
            &                            &                                                                           & \phi_M(x^{(i)})\T (Hb - b)\leq\Bmax_\Delta^{\tilde{\X}\setminus\X},
            \quad    &                            & i=1,\ldots,\tilde{N}-\hat{N},                                                                                                                                                \\
            &                            &                                                                           & c\geq 0,\,1>\eta\geq 0,\, b\in\R^{2M+1},                          &                       &
      \end{alignedat}
\end{equation}
\setlength{\abovedisplayskip}{10pt}
with $\kappa\geq\sigma_f$, $\bar{\B}\geq\norm{b}_2$, and constraint-tightening coefficients
\begin{align*}
      \hat{\eta}   & := \tfrac{2\eta + (C_{\tilde{N}}-1)\Bmin_{\tilde{N}}^{\X_0}-2A^{\tilde{\X}\setminus\X_0}_{\tilde{N}} \Bmax_{\tilde{N}}^{\tilde{\X}\setminus\X_0}}{C_{\tilde{N}}-2A^{\tilde{\X}\setminus\X_0}_{\tilde{N}}+1},
                   &                                                                                                                                                                                                                            & \hat{\gamma} := \tfrac{2 + (C_{\tilde{N}}-1)\Bmax_{\tilde{N}}^{\X_u}-2A^{\tilde{\X}\setminus\X_u}_{\tilde{N}} \Bmin_{\tilde{N}}^{\tilde{\X}\setminus\X_u}}{C_{\tilde{N}}-2A^{\tilde{\X}\setminus\X_u}_{\tilde{N}}+1}, \\
      \hat{\Delta} & := \tfrac{2(c - \varepsilon\bar{\B}\kappa) + (C_{\tilde{N}}-1)\Bmin_\Delta^\X-2A^{\tilde{\X}\setminus\X}_{\tilde{N}}\Bmax_{\Delta}^{\tilde{\X}\setminus\X}}{C_{\tilde{N}}-2A^{\tilde{\X}\setminus\X}_{\tilde{N}}+1},
                   &                                                                                                                                                                                                                            & \hat{\xi} := \tfrac{(C_{\tilde{N}}-1)\Bmax_{\tilde{N}}^{{\X}}-2A^{\tilde{\X}\setminus\X}_{\tilde{N}} \Bmin_{\tilde{N}}^{\tilde{\X}\setminus\X}}{C_{\tilde{N}}-2A^{\tilde{\X}\setminus\X}_{\tilde{N}}+1} .
\end{align*}
Here, the coefficients $C_{\tilde{N}}, A^{\tilde{\X}\setminus\X}_{\tilde{N}}, A^{\tilde{\X}\setminus\X_0}_{\tilde{N}}$, and $A^{\tilde{\X}\setminus\X_u}_{\tilde{N}}$ are obtained as follows:
\begin{align*}
    C_{{\tilde{N}}} &:= \left(1-\tfrac{2{f_{\text{max}}}}{{\tilde{Q}}}\right)^{-\frac{n}{2}},\\
    A^{\tilde{\X}\setminus\mathbb{S}}_{\tilde{N}} &:= \frac{1}{\tilde{N}} \sum_{\bar{x}\in\Theta_{\tilde{N}}\setminus\mathbb{S}} D^n_{f_{\text{max}},\tilde{Q}-f_{\text{max}}}(x-\bar{x}), \quad \mathbb{S}\in\{\X,\X_0,\X_U\},
\end{align*}
with $\tilde{N}=\tilde{Q}^n$, $f_{\text{max}}\in\N_{\geq 0}$ the maximum degree of the Fourier barrier $\B$, and $D^n_{a,b}\colon\R^n\to\R$ the Vallée-Poussin kernel
\begin{equation*}
    D^n_{a,b}(z) := \frac{1}{(b-a)^n} \prod_{i=1}^n \frac{\sin(\frac{b+a}{2}z_i)\sin(\frac{b-a}{2}z_i)}{\sin^2(\frac{z_i}{2})}.
\end{equation*}
In practice, the coefficients $A^{\tilde{\X}\setminus\X}_{\tilde{N}}, A^{\tilde{\X}\setminus\X_0}_{\tilde{N}},A^{\tilde{\X}\setminus\X_u}_{\tilde{N}}$ are computed numerically for enlarged versions of the sets $\X,\X_0,\X_U$ (see \lstinline{set_scaling} in Section~\ref{sec:config_and_run_tool}, \Circled{3}) via particle swarm optimization.
See \citet{schoenJAIR} for more details.

\section{Tool Technical Details}

\subsection{Installation}
The recommended way to install \pylucid is via the \texttt{pip} package manager, using one of the pre-built wheels,
thus avoiding the time-consuming process of compiling the software from source.
Assuming \texttt{Python>=3.8} is already present on the system, \pylucid can be installed with a single command:
\begin{lstlisting}[language=bash,numbers=none,xleftmargin=0em,backgroundcolor=\color{ipython_bg}]
pip install "pylucid[gui]" --index-url https://gitlab.com/api/v4/projects/71977529/packages/pypi/simple
\end{lstlisting}
\pylucid comes with several optional dependencies that enhance its functionality.
These can be installed by specifying the corresponding flag in square brackets after the package name.
\begin{itemize}
      \item The \texttt{gui} flag adds the \pylucid \gls{gui}.
      \item The \texttt{verification} flag adds the \dreal SMT solver, allowing for formal verification of the barriers, provided the system dynamics are known.
            Note that \dreal can only be installed on Linux and non-ARM macOS.
      \item The \texttt{gurobi} flag adds the \gurobi solver interface. Note that a valid license is required to use the \gurobi solver.
\end{itemize}
If no pre-compiled wheel is available for the desired platform, \pylucid can be installed from source by running the following commands:
\begin{lstlisting}[language=bash,numbers=none,xleftmargin=0em,backgroundcolor=\color{ipython_bg}]
git clone https://github.com/TendTo/lucid.git
cd lucid
pip install ".[gui]"
\end{lstlisting}
Building \lucid from source requires \texttt{Python>=3.8} and \texttt{Bazel>=8.0} to be installed on the system.
For more details on the installation process, please refer to the online documentation at \url{https://tendto.github.io/lucid/md_docs_2Pylucid.html}.

\subsection{Containerized Installation}

We also provide a pre-built Docker image running \pylucid in a containerized environment with all dependencies included.
To run the image with the desired configuration, run:
\begin{lstlisting}[language=bash,numbers=none,xleftmargin=0em,backgroundcolor=\color{ipython_bg}]
docker run --name lucid -it --rm \
  -v/path/to/conf.py:/config \
  ghcr.io/tendto/lucid:latest /config/conf.py
\end{lstlisting}
Alternatively, the container can also run the GUI by changing the command to
\begin{lstlisting}[language=bash,numbers=none,xleftmargin=0em,backgroundcolor=\color{ipython_bg}]
docker run --name lucid -it --rm \
  -p 3661:3661 --entrypoint pylucid-gui \
  ghcr.io/tendto/lucid:latest
\end{lstlisting}
The GUI will be accessible at \url{http://localhost:3661}.
Note that, in all cases, to use the \gurobi solver, a \gurobi Web License Service (WLS) license\footnote{See \url{https://www.gurobi.com/features/web-license-service/}} will have to be mounted in the container.

\subsection{Configuration Formats}
\pylucid configuration can be stored in a \texttt{.yaml} file (recommended), \texttt{.json} file,
generated via a Python script producing a \texttt{Configuration} object, or provided directly via command line arguments.
Moreover, system transition data can be either embedded directly in the configuration file or provided separately in the form of \texttt{.csv}, \texttt{.mat}, \texttt{.npy}, or \texttt{.npz} files.
For example, the following configurations are equivalent:

\begin{minipage}{.45\columnwidth}
\begin{lstlisting}[language=yaml,style={bgnonumbers},caption={YAML configuration.},captionpos=b,label={lst:configuration.yaml},backgroundcolor=\color{ipython_bg}]
seed: 42
X_bounds: "RectSet([-1], [1])"
X_init: "RectSet([-0.5], [0.5])"
X_unsafe:
  - "RectSet([-1], [-0.9])"
  - "RectSet([0.9], [1])"
x_samples: [[1.0], [-1.0]]
xp_samples: [[0.5], [-0.5]]
lambda: 1.0e-5
time_horizon: 15
sigma_f: 18.0
sigma_l: 0.034
num_frequencies: 5
lattice_resolution: 700
\end{lstlisting}
\end{minipage}\hfill
\begin{minipage}{.45\columnwidth}
\begin{lstlisting}[language=json,style={bgnonumbers},caption={JSON configuration.},captionpos=b,label={lst:configuration.json}]
{
  "seed": 42,
  "X_bounds": "RectSet([-1], [1])",
  "X_init": "RectSet([-0.5], [0.5])",
  "X_unsafe": [
    "RectSet([-1], [-0.9])", 
    "RectSet([0.9], [1])"
  ],
  "x_samples": [[1], [-1]],
  "xp_samples": [[0.5], [-0.5]],
  "lambda": 0.00001,
  "time_horizon": 15,
  "sigma_f": 18,
  "sigma_l": 0.034,
  "num_frequencies": 5,
  "lattice_resolution": 700
}
\end{lstlisting}
\end{minipage}
\begin{minipage}{.45\columnwidth}
\begin{lstlisting}[language=iPython,style={bgnonumbers},caption={Python configuration.},captionpos=b,label={lst:configuration.py}]
from pylucid import *
import numpy as np

def scenario_config() -> Configuration:
  c = Configuration(seed=42, lambda_=1e-5)
  c.X_bounds = RectSet([(-1, 1)])
  c.X_init = RectSet([(-0.5, 0.5)])
  c.X_unsafe = MultiSet(
      RectSet([(-1, -0.9)]), 
      RectSet([(0.9, 1)]),
  )
  c.x_samples = np.array([[1.0], [-1.0]])
  c.xp_samples = np.array([[0.5], [-0.5]])
  c.time_horizon = 15
  c.sigma_f = 18.0
  c.sigma_l = 0.034
  c.num_frequencies = 5
  c.lattice_resolution = 700
  return c
\end{lstlisting}
\end{minipage}\hfill
\begin{minipage}{.45\columnwidth}
\begin{lstlisting}[language=bash,style={bgnonumbers},caption={Command line configuration assuming that the transition samples have been stored as \texttt{.csv} files.},captionpos=b,label={lst:configuration.sh}]
pylucid \ 
  --seed 42 \
  --X_bounds "RectSet([-1], [1])" \
  --X_init "RectSet([-0.5], [0.5])" \
  --X_unsafe "MultiSet([RectSet([-1], [-0.9]),
  RectSet([0.9], [1])])" \
  --lambda "1.0e-5" \
  --time_horizon 15 \
  --sigma_f 18.0 \
  --sigma_l 0.034 \
  --num_frequencies 5 \
  --lattice_resolution 700 \
  -i x_samples.csv \
  -o xp_samples.csv
\end{lstlisting}
\end{minipage}

\subsection{Graphical User Interface}

\pylucid provides a \gls{gui} to guide the user in the creation of the scenario configuration and presenting the results in an intuitive format.
Running \lstinline|pylucid-gui| will open a browser tab with the interface shown in Figure~\ref{fig:gui-full},
while a local server will listen for requests coming from the \gls{gui}, computing and returning the results.
Following the same enumeration as the arrows in the figure, we provide a list of the main components and features of the \gls{gui}:

\begin{figure}
      \centering
      \shadowimage[width=16.5cm]{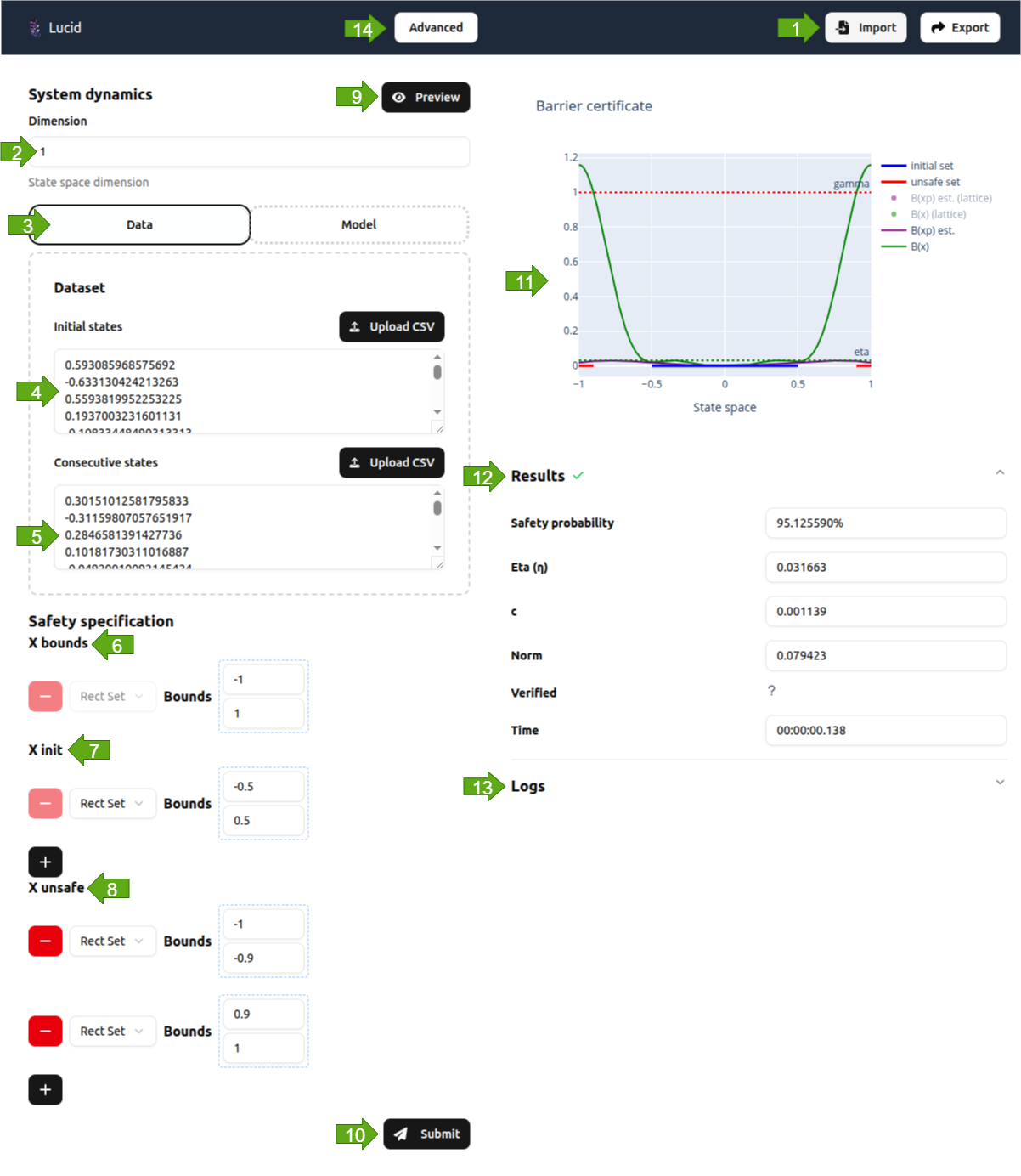}
      \caption{
            \pylucid's \gls{gui} as presented in the browser.
            The left half of the page allows the user to provide the transition samples and the safety specification to verify.
            The right half of the page shows the results produced by \lucid, including a plot of the computed barrier function and detailed logs.}
      \label{fig:gui-full}
\end{figure}

\begin{enumerate}
      \item It is easy to import or export configurations in JSON format.
            In the \textit{Import} menu, the user can paste the configuration or load it from a file.
            The experiments discussed in this paper are also available and can be selected from a dropdown list.
            The \textit{Export} menu can be used to copy the current configuration to the clipboard or download it locally.
      \item The dimension $n$ of the system is specified. Here, $n=1$.
      \item The user can provide transition samples directly by copy--pasting them into the text fields in the Data tab.
            Alternatively, if a model describing the system’s expected behavior is available, the user can specify it in the Model tab.
            Lucid will then automatically generate a set of transitions from the model to be used in the computation of the barrier function.
            Note that this model-based generation is mainly intended for reference and convenience.
      \item The $x^1,\ldots,x^N$ samples are specified in CSV format. They can be edited manually or loaded from a file.
      \item The $x^1_+,\ldots,x^N_+$ samples are specified analogously.
      \item The state space $\X$ is defined by a \texttt{RectSet}. Here, $\X=[-1,1]$.
      \item The initial set $\X_0$ is defined from one of the available type options. More sets can be added via the \textit{+} button. Here, $\X_0=[-0.5,0.5]$.
      \item The unsafe set $\X_U$ is defined analogously. Here, $\X_U=[-1, -0.9] \cup [0.9, 1]$.
      \item The \textit{Preview} button visualizes the expected stochastic behavior of the black-box system, predicted from the transition samples or alternatively as specified in the \textit{Model} tab,
            alongside the bounding boxes of the state space $\X$, the initial set $\X_0$, and the unsafe set $\X_U$.
      \item The \textit{Submit} button starts the computation of the barrier function with the current configuration.
      \item The plot of the transition function or the computed barrier function.
      \item The numerical results of the computation, including the lower bound on the safety probability, values of $\eta$ and $c$, and the computation time.
            If a model of the expected system behavior is provided, it is also possible to formally verify the computed barrier function via \dreal.
      \item The logs provide information on the progress of the computation, updated in real-time.
            The verbosity of the logs can be adjusted in the \textit{Advanced} options.
      \item The main \gls{gui} displays the most important configuration options.
            More advanced options can be accessed via the \textit{Advanced} button.
            This includes the kernel \hp, time horizon, number of frequencies and lattice size, \gls{lp} solver selection, etc.
\end{enumerate}

\section{Benchmarks}
\label{app:benchmarks}
We provide a more detailed description of the benchmarks presented in Section~\ref{sec:experiments}, including the corresponding configuration files and an extended list of experiments.
All benchmarks were run on a Windows 10 machine with an AMD Ryzen 9 5950X 16-Core Processor @ 3.40 GHz, NVIDIA GeForce RTX 3090 GPU, and 64 GB of RAM.
To ensure reproducibility, all runs had the random seed set to $42$.
We ascertained that the variance of the safety probability is low across different seeds, and can be reduced further by increasing the number of samples used to fit the \estimator.
The scripts for running the benchmarks and tuning the \hp are available in the \texttt{benchmarks/integration/} folder of the repository.
Note that \texttt{Python>=3.9} is required to run the scripts, as some dependencies used to track the benchmarks' metrics across multiple runs are not compatible with \texttt{Python 3.8}.
The hyperparameter tuning of the \estimator was performed with the \texttt{LbfgsTuner}, bounding the value of $\sigma_l$ between $[10^{-5}, 10^5]$, and refined with the \texttt{GridSearchTuner}.
For additional flexibility, we use individual lengthscales $\sigma_l$ and $\sigma_{l_f}$ for the input kernel and output kernel, respectively.
We use $N=1000$ sample transitions and the \gurobi solver for all experiments;
Other optimizers, such as \alglib or \highs, can be used instead, but may yield different results, especially in terms of performance.

\subsection{Linear}

We consider a black-box system with the following dynamics:
\begin{equation*}
      {x}_{t+1} = 0.5 {x}_{t} + w_t,
\end{equation*}
where $w_t\sim\mathcal{N}(\cdotx\vert 0,0.01I_1)$.
The data sampled from this black-box system is used as input for \lucid.
Given $\X = [-1, 1]$, $\X_0 = [-0.5, 0.5]$, and $\X_U = [-1, -0.9] \cup [0.9, 1]$,
we want to ensure that the system, starting in $\X_0$, does not enter the unsafe regions $\X_U$ within $T=15$ time steps.
We set the kernel \hp to $\sigma_f=1$ and $\lambda=0.00001$.
The complete configuration for the linear example benchmark is shown in Listing~\ref{lst:linear}, with a satisfying barrier plotted in Figure~\ref{fig:example}.
\begin{lstlisting}[language=yaml,style={bgnonumbers},caption={Configuration for the linear example.},captionpos=b,label={lst:linear},backgroundcolor=\color{ipython_bg}]
b_norm: 2.0
estimator: KernelRidgeRegressor
feature_map: LinearTruncatedFourierFeatureMap
feature_sigma_l: 0.0925
gamma: 1.0
kernel: GaussianKernel
lambda: 1.0e-5
lattice_resolution: 300
noise_scale: 0.01
num_frequencies: 6
num_samples: 1000
optimiser: GurobiOptimiser
plot: true
seed: 42
set_scaling: 0.04000000000000001
sigma_f: 1.0
sigma_l: 0.04465750366442458
system_dynamics: [x1 / 2]
time_horizon: 15
verbose: 3
X_bounds:
  - RectSet:
      - [-1, 1]
X_init:
  - RectSet:
      - [-0.5, 0.5]
X_unsafe:
  - RectSet:
      - [-1, -0.9]
  - RectSet:
      - [0.9, 1]
\end{lstlisting}

\begin{table}[tb]
      \centering
      \begin{tabular}{ccccccccc}
            \toprule
            \textbf{\#Freq.} & \textbf{Lattice} & $\sigma_{l_f}$ & $\sigma_l$ & \textbf{Set}   & $\eta$ & $c$  & \textbf{Runtime} & \textbf{Safety} \\
                           & \textbf{Size}    &                &            & \textbf{Scaling} &        &      & [mm:ss]          & \textbf{Prob.}  \\
            \midrule
            6              & 300              & [0.09]         & [0.044]    & 4\%            & 0.01   & 0.00 & 0:01             & 93.07\%         \\
            7              & 400              & [0.09]         & [0.197]    & 3\%            & 0.02   & 0.00 & 0:01             & 92.99\%         \\
            6              & 400              & [0.09]         & [0.131]    & 4\%            & 0.02   & 0.00 & 0:01             & 91.61\%         \\
            5              & 400              & [0.13]         & [0.035]    & 5\%            & 0.04   & 0.00 & 0:01             & 89.13\%         \\
            6              & 300              & [0.13]         & [0.345]    & 3\%            & 0.05   & 0.00 & 0:01             & 88.69\%         \\
            8              & 400              & [0.09]         & [0.167]    & 3\%            & 0.02   & 0.01 & 0:01             & 85.85\%         \\
            7              & 400              & [0.09]         & [0.123]    & 4\%            & 0.02   & 0.01 & 0:01             & 84.94\%         \\
            6              & 400              & [0.09]         & [0.024]    & 5\%            & 0.02   & 0.01 & 0:01             & 83.91\%         \\
            8              & 300              & [0.09]         & [0.027]    & 3\%            & 0.05   & 0.01 & 0:01             & 79.36\%         \\
            4              & 400              & [0.18]         & [0.081]    & 6\%            & 0.12   & 0.01 & 0:01             & 74.87\%         \\
            \bottomrule
      \end{tabular}

      \caption{
            \linear results, sorted by the last column. For each combination of number of frequencies, $M$, and lattice size (i.e., number of lattice points per dimension), we report the values of the $\sigma_l$ used in the estimator and the feature map, how much sets were scaled w.r.t the periodic space, $\eta$, $c$, the runtime, and the achieved lower bound on the safety probability.}
      \label{tab:results-linear}
\end{table}

\subsection{Barrier 2}

We consider a black-box system with the following nonlinear dynamics:
\begin{equation*}
      \begin{bmatrix}
            {x}_{1, t+1} \\
            {x}_{2, t+1}
      \end{bmatrix}
      = \begin{bmatrix}
            {x}_{1, t} \\
            {x}_{2, t}
      \end{bmatrix} + 0.1 \begin{bmatrix}
            {x}_{2, t} - 1 + e^{-x_{1, t}} \\
            -\sin^2(x_{1, t})
      \end{bmatrix} + w_t,
\end{equation*}
where $w_t\sim\mathcal{N}(\cdotx\vert 0,0.01I_2)$.
The data sampled from this black-box system is used as input for \lucid.
Given
\begin{align*}
       & \X = [ -2, 2 ] \times [ -2 , 2 ]   ,                              \\
       & \X_0 = \{ [x_1, x_2] : (x_1 + 0.5)^2 + (x_2 - 0.5)^2 \leq 0.4 \}, \\
       & \X_U = \{ [x_1, x_2] : (x_1 - 0.7)^2 + (x_2 + 0.7)^2 \leq 0.3 \},
\end{align*}
we want to ensure that the system, starting in $\X_0$, does not enter the unsafe regions $\X_U$ within $T=5$ time steps.
The system dynamics and safety specification are visualized in Figure~\ref{fig:model-barrier2}.
The kernel \hp are set to $\sigma_f=1$ and $\lambda=\smash{10^{-6}}$.
The complete configuration for the \barrII benchmark is shown in Listing~\ref{lst:barrier2}, with a satisfying barrier is plotted in Figure~\ref{fig:CSBarr2}.
A list of experiments using different combinations of frequencies and $\sigma_l$ values, showing their impact on performance and final result, is presented in Table~\ref{tab:results-barrier2}.
\begin{lstlisting}[language=yaml,style={bgnonumbers},caption={Configuration for \barrII.},captionpos=b,label={lst:barrier2}]
b_norm: 16.0
estimator: KernelRidgeRegressor
feature_map: LinearTruncatedFourierFeatureMap
feature_sigma_l: [0.16000000000000003, 0.11677777777777779]
gamma: 1.0
kernel: GaussianKernel
lambda: 1.0e-06
lattice_resolution: 330
noise_scale: 0.01
num_frequencies: 7
num_samples: 1000
optimiser: GurobiOptimiser
plot: true
seed: 42
set_scaling: 0.05
sigma_f: 1.0
sigma_l: [0.40501051998808607, 0.3580734352248586]
system_dynamics:
  - "x1 + 0.1 * (x2 - 1 + exp(-x1))"
  - "x2 + 0.1 * -(sin(x1)^2)"
time_horizon: 5
verbose: 3
X_bounds:
  - RectSet:
      - [-2, 2]
      - [-2, 2]
X_init:
  - SphereSet:
      center: [-0.5, 0.5]
      radius: 0.4
X_unsafe:
  - SphereSet:
      center: [0.7, -0.7]
      radius: 0.3
\end{lstlisting}

\begin{figure}[ht]
      \centering
      \input{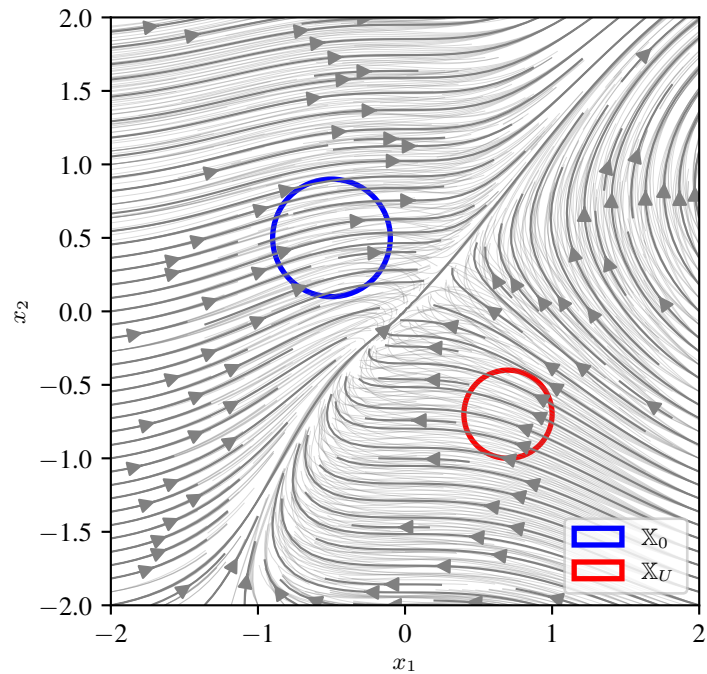}
      \caption{Visualization of the stochastic system behavior (based on 10 random transitions from an $100\times 100$ grid of initial states) and safety specification for the \barrII benchmark.}
      \label{fig:model-barrier2}
\end{figure}

\begin{table}[tb]
      \centering
      \begin{tabular}{ccccccccc}
            \toprule
            \textbf{\#Freq.} & \textbf{Lattice} & $\sigma_{l_f}$ & $\sigma_l$   & \textbf{Set}   & $\eta$ & $c$  & \textbf{Runtime} & \textbf{Safety} \\
                           & \textbf{Size}    &                &              & \textbf{Scaling} &        &      & [mm:ss]          & \textbf{Prob.}  \\
            \midrule
            7              & 330              & [0.16, 0.12]   & [0.41, 0.36] & 5\%            & 0.18   & 0.04 & 4:38             & 63.17\%         \\
            7              & 330              & [0.12, 0.15]   & [0.45, 0.34] & 5\%            & 0.16   & 0.05 & 3:21             & 59.17\%         \\
            7              & 330              & [0.16, 0.13]   & [0.52, 0.44] & 5\%            & 0.19   & 0.05 & 4:18             & 58.38\%         \\
            7              & 330              & [0.20, 0.10]   & [0.60, 0.32] & 5\%            & 0.20   & 0.05 & 4:00             & 54.82\%         \\
            7              & 330              & [0.20, 0.12]   & [0.68, 0.41] & 5\%            & 0.22   & 0.05 & 4:13             & 54.11\%         \\
            7              & 330              & [0.16, 0.15]   & [0.30, 0.42] & 5\%            & 0.20   & 0.05 & 4:07             & 53.88\%         \\
            7              & 330              & [0.24, 0.10]   & [0.72, 0.21] & 5\%            & 0.23   & 0.05 & 4:21             & 53.46\%         \\
            6              & 330              & [0.40, 0.07]   & [0.75, 0.21] & 5\%            & 0.25   & 0.04 & 2:28             & 53.13\%         \\
            6              & 330              & [0.20, 0.13]   & [0.68, 0.26] & 5\%            & 0.21   & 0.05 & 1:39             & 52.36\%         \\
            7              & 330              & [0.24, 0.08]   & [0.44, 0.25] & 5\%            & 0.22   & 0.05 & 4:28             & 52.31\%         \\
            7              & 330              & [0.24, 0.15]   & [0.72, 0.35] & 4\%            & 0.26   & 0.05 & 6:27             & 50.38\%         \\
            6              & 330              & [0.36, 0.08]   & [0.62, 0.25] & 5\%            & 0.25   & 0.05 & 2:40             & 49.13\%         \\
            7              & 330              & [0.20, 0.13]   & [0.58, 0.31] & 5\%            & 0.24   & 0.05 & 4:41             & 48.91\%         \\
            6              & 330              & [0.24, 0.15]   & [0.66, 0.36] & 5\%            & 0.27   & 0.05 & 2:03             & 48.54\%         \\
            6              & 330              & [0.24, 0.12]   & [0.81, 0.33] & 5\%            & 0.24   & 0.06 & 1:57             & 47.84\%         \\
            7              & 330              & [0.20, 0.15]   & [0.63, 0.38] & 5\%            & 0.27   & 0.05 & 4:24             & 47.64\%         \\
            6              & 330              & [0.32, 0.12]   & [0.65, 0.34] & 5\%            & 0.27   & 0.05 & 2:19             & 46.55\%         \\
            6              & 330              & [0.40, 0.08]   & [0.55, 0.29] & 5\%            & 0.26   & 0.06 & 2:25             & 46.46\%         \\
            6              & 330              & [0.28, 0.12]   & [0.77, 0.30] & 5\%            & 0.28   & 0.05 & 1:59             & 46.22\%         \\
            6              & 330              & [0.28, 0.13]   & [0.73, 0.31] & 5\%            & 0.30   & 0.05 & 1:43             & 44.72\%         \\
            6              & 330              & [0.32, 0.13]   & [0.84, 0.47] & 5\%            & 0.30   & 0.05 & 2:14             & 44.69\%         \\
            \bottomrule
      \end{tabular}
      \caption{
            \barrII results, sorted by the last column. For each combination of number of frequencies, $M$, and lattice size (i.e., number of lattice points per dimension), we report the values of $c$, $\lambda$, the runtime, and the achieved lower bound on the safety probability.}
      \label{tab:results-barrier2}
\end{table}

\subsection{Barrier 3}

We consider a black-box system that has the following dynamics:
\begin{equation*}
      \begin{bmatrix}
            {x}_{1, t+1} \\
            {x}_{2, t+1}
      \end{bmatrix}
      = \begin{bmatrix}
            {x}_{1, t} \\
            {x}_{2, t}
      \end{bmatrix} + 0.1 \begin{bmatrix}
            {x}_{2, t} \\
            \frac{1}{3} {x}^3_{1, t} - {x}_{1,t} - {x}_{2,t}
      \end{bmatrix} + w_t,
\end{equation*}
where $w_t\sim\mathcal{N}(\cdotx\vert 0,0.01I_2)$.
The data sampled from this black-box system is used as input for \lucid.
Given
\begin{align*}
       & \X = [ -3, 2.5 ] \times [ -2 , 1 ]                                                 \\
       & \X_0 = [ 1 , 2 ] \times [ -0.7 , 0.3 ] \cup [ -1.8 , -1.4 ] \times [ -0.1 , 0.1 ]  \\
       & \qquad \cup [-1.4, -1.2] \times [-0.5 , 0.1]                                       \\
       & \X_U = [ 0.4 , 0.6 ] \times [ 0.2 , 0.6 ] \cup [ 0.6 , 0.7 ] \times [ 0.2 , 0.4 ],
\end{align*}
we want to ensure that the system, starting in $\X_0$, does not enter the unsafe regions $\X_U$ within $T=5$ time steps.
The system dynamics and safety specification can be visualized in Figure~\ref{fig:model-barrier3}.
The kernel \hp were set to $\sigma_f=1$ and $\lambda=10^{-8}$.
The complete configuration for the \barrIII benchmark is shown in Listing~\ref{lst:barrier3}, while the barrier function is plotted in Figure~\ref{fig:CSBarr3}.
A list of experiments using different combinations of frequencies and $\sigma_l$ values, showing their impact on performance and final result, is presented in Table~\ref{tab:results-barrier3}.
\begin{lstlisting}[language=yaml,style={bgnonumbers},caption={Configuration for \barrIII.},captionpos=b,label={lst:barrier3}]
b_norm: 7.0
estimator: KernelRidgeRegressor
feature_map: LinearTruncatedFourierFeatureMap
feature_sigma_l: [0.05, 0.15]
gamma: 1.0
kernel: GaussianKernel
lambda: 1.e-8
lattice_resolution: 330
noise_scale: 0.01
num_frequencies: 7
num_samples: 1000
optimiser: GurobiOptimiser
plot: true
seed: 42
set_scaling: 0.025
sigma_f: 1.0
sigma_l: [0.320588187057447, 0.15721319058133434]
system_dynamics:
  - x1 + 0.1 * x2
  - x2 + 0.1 * (-x1 - x2 + 1 / 3 * x1 ^ 3)
time_horizon: 5
verbose: 3
X_bounds: RectSet([-3, -2], [2.5,  1])
X_init:
  - RectSet([1, -0.7], [2,  0.3])
  - RectSet([-1.8, -0.1], [-1.4,  0.1])
  - RectSet([-1.4, -0.5], [-1.2,  0.1])
X_unsafe:
  - RectSet([0.4, 0.2], [0.6, 0.6])
  - RectSet([0.6, 0.2], [0.7, 0.4])
\end{lstlisting}

\begin{figure}[ht]
      \centering
      \input{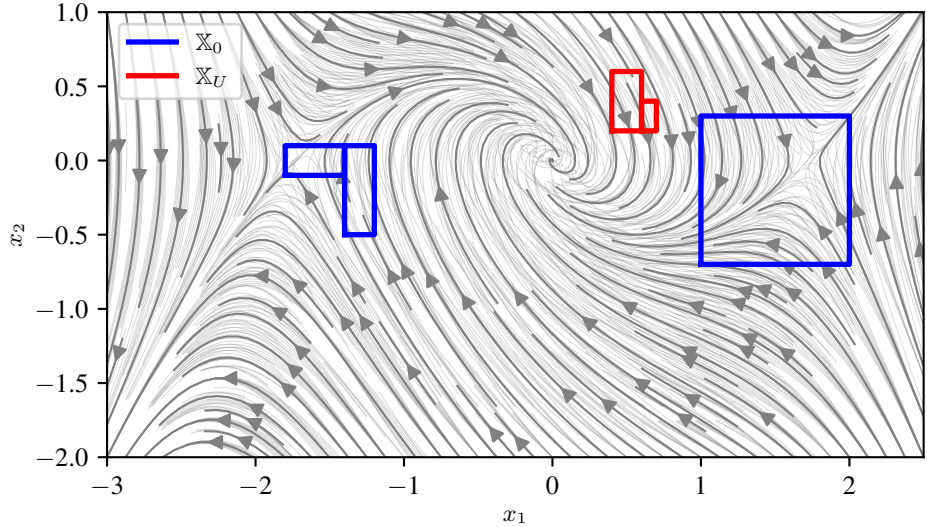}
      \caption{Visualization of the stochastic system behavior (based on 10 random transitions from an $100\times 100$ grid of initial states) and safety specification for the \barrIII benchmark.}
      \label{fig:model-barrier3}
\end{figure}

\begin{table}[tb]
      \centering
      \begin{tabular}{ccccccccc}
            \toprule
            \textbf{\#Freq.} & \textbf{Lattice} & $\sigma_{l_f}$ & $\sigma_l$   & \textbf{Set}   & $\eta$ & $c$  & \textbf{Runtime} & \textbf{Safety} \\
                           & \textbf{Size}    &                &              & \textbf{Scaling} &        &      & [mm:ss]          & \textbf{Prob.}  \\
            \midrule
            7              & 330              & [0.05, 0.15]   & [0.32, 0.16] & 2\%            & 0.29   & 0.04 & 3:20             & 51.63\%         \\
            7              & 330              & [0.05, 0.20]   & [0.29, 0.57] & 2\%            & 0.29   & 0.04 & 3:29             & 50.92\%         \\
            7              & 330              & [0.05, 0.25]   & [0.30, 0.23] & 2\%            & 0.29   & 0.04 & 3:52             & 50.91\%         \\
            7              & 330              & [0.05, 0.25]   & [0.19, 0.71] & 2\%            & 0.29   & 0.04 & 3:16             & 50.50\%         \\
            7              & 330              & [0.05, 0.20]   & [0.26, 0.81] & 2\%            & 0.29   & 0.04 & 3:24             & 49.63\%         \\
            7              & 330              & [0.05, 0.15]   & [0.15, 0.62] & 2\%            & 0.28   & 0.04 & 2:43             & 49.15\%         \\
            7              & 330              & [0.05, 0.30]   & [0.20, 0.77] & 2\%            & 0.30   & 0.04 & 3:20             & 48.91\%         \\
            7              & 330              & [0.05, 0.25]   & [0.19, 0.21] & 2\%            & 0.30   & 0.04 & 3:13             & 48.72\%         \\
            7              & 330              & [0.05, 0.30]   & [0.36, 1.27] & 2\%            & 0.30   & 0.04 & 3:40             & 47.94\%         \\
            7              & 330              & [0.05, 0.30]   & [0.20, 0.83] & 2\%            & 0.30   & 0.04 & 3:23             & 47.47\%         \\
            7              & 330              & [0.05, 0.35]   & [0.29, 0.39] & 2\%            & 0.31   & 0.04 & 3:30             & 47.46\%         \\
            7              & 330              & [0.05, 0.40]   & [0.24, 1.25] & 2\%            & 0.30   & 0.05 & 3:23             & 45.74\%         \\
            7              & 330              & [0.05, 0.20]   & [0.26, 0.70] & 2\%            & 0.32   & 0.04 & 2:46             & 45.40\%         \\
            7              & 330              & [0.05, 0.35]   & [0.19, 0.94] & 2\%            & 0.32   & 0.05 & 3:47             & 45.10\%         \\
            7              & 330              & [0.05, 0.15]   & [0.21, 0.38] & 2\%            & 0.32   & 0.05 & 2:45             & 44.36\%         \\
            7              & 330              & [0.05, 0.40]   & [0.22, 0.97] & 2\%            & 0.33   & 0.05 & 3:17             & 44.14\%         \\
            7              & 330              & [0.05, 0.35]   & [0.10, 1.10] & 2\%            & 0.31   & 0.05 & 2:55             & 43.64\%         \\
            7              & 330              & [0.05, 0.40]   & [0.22, 0.92] & 2\%            & 0.33   & 0.05 & 2:18             & 42.42\%         \\
            7              & 330              & [0.05, 0.45]   & [0.18, 0.81] & 2\%            & 0.34   & 0.05 & 0:48             & 41.21\%         \\
            8              & 330              & [0.05, 0.10]   & [0.20, 0.36] & 2\%            & 0.30   & 0.06 & 5:21             & 41.03\%         \\
            8              & 330              & [0.05, 0.15]   & [0.18, 0.38] & 2\%            & 0.32   & 0.05 & 2:52             & 40.99\%         \\
            8              & 330              & [0.05, 0.10]   & [0.27, 0.44] & 2\%            & 0.32   & 0.05 & 4:20             & 40.25\%         \\
            7              & 330              & [0.05, 0.45]   & [0.07, 1.38] & 2\%            & 0.32   & 0.06 & 1:04             & 39.90\%         \\
            8              & 330              & [0.05, 0.15]   & [0.29, 0.14] & 2\%            & 0.34   & 0.05 & 2:30             & 39.42\%         \\
            7              & 330              & [0.05, 0.45]   & [0.24, 0.28] & 2\%            & 0.35   & 0.05 & 2:57             & 38.76\%         \\
            7              & 330              & [0.05, 0.10]   & [0.33, 0.29] & 2\%            & 0.36   & 0.05 & 2:28             & 38.73\%         \\
            8              & 330              & [0.05, 0.20]   & [0.31, 0.49] & 2\%            & 0.34   & 0.06 & 2:08             & 38.52\%         \\
            7              & 330              & [0.05, 0.10]   & [0.36, 0.27] & 2\%            & 0.38   & 0.05 & 2:35             & 38.39\%         \\
            8              & 330              & [0.05, 0.10]   & [0.21, 0.19] & 2\%            & 0.33   & 0.06 & 5:09             & 38.07\%         \\
            6              & 330              & [0.05, 0.20]   & [0.16, 0.53] & 2\%            & 0.26   & 0.07 & 1:35             & 37.66\%         \\
            7              & 330              & [0.05, 0.50]   & [0.17, 0.70] & 2\%            & 0.35   & 0.06 & 0:56             & 36.97\%         \\
            8              & 330              & [0.05, 0.20]   & [0.22, 0.69] & 2\%            & 0.34   & 0.06 & 2:51             & 36.87\%         \\
            8              & 330              & [0.05, 0.25]   & [0.26, 0.44] & 2\%            & 0.33   & 0.06 & 2:30             & 36.64\%         \\
            7              & 330              & [0.05, 0.55]   & [0.25, 0.93] & 2\%            & 0.35   & 0.06 & 1:02             & 36.49\%         \\
            7              & 330              & [0.05, 0.10]   & [0.08, 0.31] & 2\%            & 0.38   & 0.05 & 2:27             & 36.43\%         \\
            8              & 330              & [0.05, 0.15]   & [0.22, 0.26] & 2\%            & 0.35   & 0.06 & 2:03             & 35.55\%         \\
            7              & 330              & [0.05, 0.50]   & [0.18, 0.78] & 2\%            & 0.36   & 0.06 & 1:18             & 35.19\%         \\
            7              & 330              & [0.05, 0.55]   & [0.19, 1.07] & 2\%            & 0.36   & 0.06 & 1:19             & 34.39\%         \\
            8              & 330              & [0.05, 0.25]   & [0.18, 0.69] & 2\%            & 0.34   & 0.06 & 2:46             & 33.75\%         \\
            6              & 330              & [0.05, 0.30]   & [0.19, 0.67] & 2\%            & 0.34   & 0.07 & 1:57             & 30.01\%         \\
            6              & 330              & [0.05, 0.30]   & [0.19, 0.81] & 2\%            & 0.34   & 0.08 & 1:32             & 27.17\%         \\
            \bottomrule
      \end{tabular}
      \caption{\barrIII results, sorted by the last column. For each combination of number of frequencies $M$ and lattice size (i.e., number of lattice points per dimension), we report the values of $c$, $\lambda$, the runtime, and the achieved lower bound on the safety probability.}
      \label{tab:results-barrier3}
\end{table}

\subsection{Overtaking}

We consider a scenario where an autonomous vehicle controlled by a \gls{nn} is overtaking another vehicle.
The dynamics of the ego vehicle are given by Dubin's car model with additive noise $w = \smash{\begin{bmatrix}w_t^1 & w_t^2 & w_t^3\end{bmatrix}\T}$ where each component $w_t^i$ is drawn from a zero-mean Gaussian with standard deviation $0.01$, $0.01$, and $0.001$, respectively.
The steering wheel angle is supplied by the \gls{nn} controller and we travel at a fixed velocity.
Given $N=1000$ sample transitions and the sets
\begin{align*}
       & \X = [ 1, 90 ] \times [ -7, 19 ] \times [ -\pi, \pi ],
       &                                                        & \X_0 = [ 1, 3 ] \times [ -1, 1 ] \times [ -0.5, 0.5 ],
       &                                                        & \X_U = [ 1, 90 ] \times [ -7, -6 ] \times [ -\pi, \pi ]                                                                       \\
       &                                                        &                                                         &  &  & \qquad \cup [ 1, 90 ] \times [ 18, 19 ] \times [ -\pi, \pi ]  \\
       &                                                        &                                                         &  &  & \qquad \cup [ 40, 45 ] \times [ -6, 6 ] \times [ -\pi, \pi ],
\end{align*}
we want to ensure that the system, starting in $\X_0$, does not enter the unsafe regions $\X_U$ within $T=5$ time steps.
The sampled transitions and safety specification are visualized in Figure~\ref{fig:model-overtaking}.
The kernel \hp are set to $\sigma_f=7$, and $\lambda=10^{-5}$.
The complete configuration for the \overtaking benchmark is shown in Listing~\ref{lst:overtaking}.
A list of experiments using different combinations of frequencies, lattice sizes and $\sigma_l$ values, showing their impact on performance and final result, is presented in Table~\ref{tab:results-overtaking}.
\begin{lstlisting}[language=yaml,style={bgnonumbers},caption={Configuration for \overtaking. The transition samples were generated by a separate script and appended to the configuration, abbreviated here for conciseness.},captionpos=b,label={lst:overtaking}]
b_norm: 5.0
estimator: KernelRidgeRegressor
feature_map: LinearTruncatedFourierFeatureMap
feature_sigma_l: [0.525, 0.05, 0.525]
gamma: 1.0
kernel: GaussianKernel
lambda: 1.0e-05
lattice_resolution: 70
num_frequencies: 5
optimiser: GurobiOptimiser
plot: true
seed: 42
set_scaling: 0.03
sigma_f: 7.0
sigma_l: [0.05414576430336046, 0.09360528166417184, 5.078332038463804]
time_horizon: 5
verbose: 4
X_bounds:
  - RectSet: [[1, 90], [-7, 19], [-3.14159265358979, 3.14159265358979]]
X_init:
  - RectSet: [[1, 3], [-1, 1], [-0.5, 0.5]]
X_unsafe:
  - RectSet: [[1, 90], [-7, -6], [-3.14159265358979, 3.14159265358979]]
  - RectSet: [[1, 90], [18, 19], [-3.14159265358979, 3.14159265358979]]
  - RectSet: [[40, 45], [-6, 6], [-3.14159265358979, 3.14159265358979]]
x_samples:
  - [..., ..., ...]
  - ...
  - [..., ..., ...]
xp_samples:
  - [..., ..., ...]
  - ...
  - [..., ..., ...]
\end{lstlisting}

\begin{figure}[ht]
      \centering
      \input{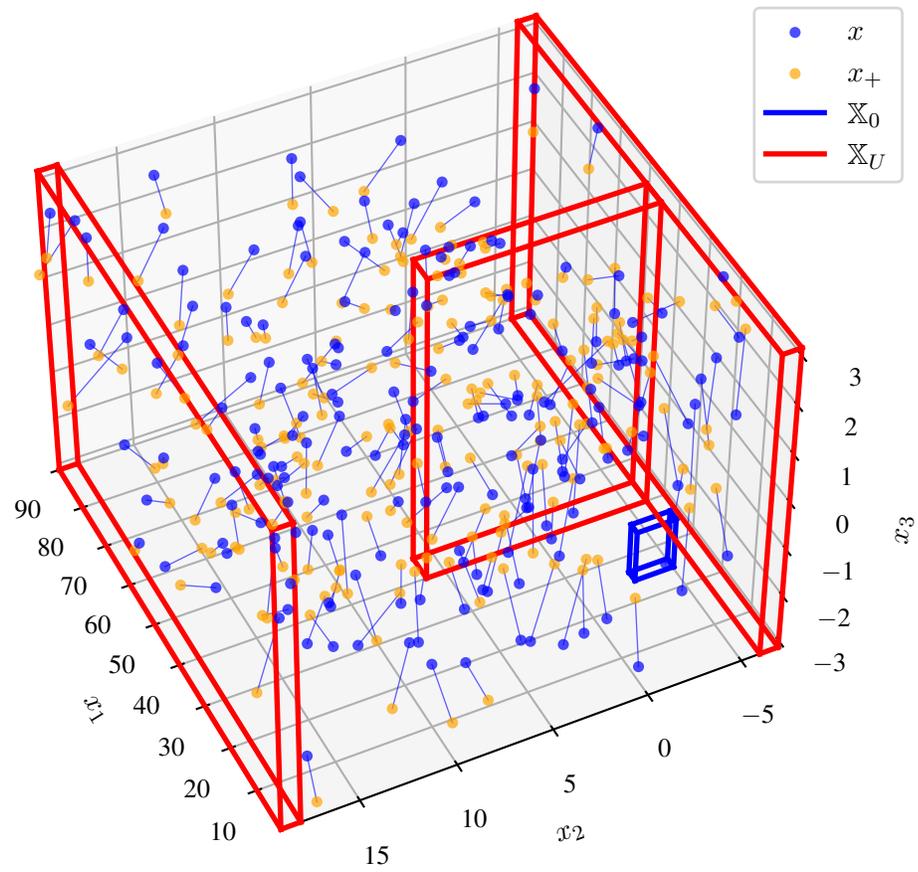}
      \caption{Visualization of $N=200$ sample transitions and safety specification for the \overtaking benchmark.}
      \label{fig:model-overtaking}
\end{figure}

\begin{table}[tb]
      \centering
      \begin{tabular}{ccccccccc}
            \toprule
            \textbf{\#Freq.} & \textbf{Lattice} & $\sigma_{l_f}$     & $\sigma_l$          & \textbf{Set}   & $\eta$ & $c$  & \textbf{Runtime} & \textbf{Safety} \\
                           & \textbf{Size}    &                    &                     & \textbf{Scaling} &        &      & [mm:ss]          & \textbf{Prob.}  \\
            \midrule
            5              & 70               & [0.53, 0.05, 0.53] & [0.05, 0.09, 5.08]  & 3\%            & 0.46   & 0.00 & 93:56            & 53.66\%         \\
            5              & 70               & [0.53, 0.05, 1.00] & [0.08, 0.37, 0.29]  & 5\%            & 0.48   & 0.00 & 94:03            & 52.25\%         \\
            5              & 70               & [0.53, 0.05, 1.00] & [0.11, 0.02, 0.43]  & 3\%            & 0.50   & 0.00 & 75:01            & 50.20\%         \\
            5              & 70               & [0.53, 0.05, 0.53] & [1.70, 0.01, 0.71]  & 4\%            & 0.57   & 0.00 & 96:02            & 43.15\%         \\
            5              & 70               & [0.53, 0.05, 0.53] & [17.70, 0.09, 0.01] & 5\%            & 0.57   & 0.00 & 103:49           & 42.58\%         \\
            5              & 80               & [0.53, 0.05, 0.53] & [0.00, 7227, 1.02]  & 2\%            & 0.61   & 0.00 & 111:55           & 38.81\%         \\
            5              & 70               & [0.53, 0.05, 0.05] & [3.24, 0.07, 0.02]  & 4\%            & 0.77   & 0.00 & 37:30            & 22.63\%         \\
            5              & 70               & [0.53, 0.05, 0.05] & [3.24, 0.06, 0.07]  & 3\%            & 0.78   & 0.00 & 44:36            & 22.08\%         \\
            5              & 70               & [0.53, 0.05, 1.00] & [2.05, 0.32, 0.03]  & 4\%            & 0.83   & 0.00 & 67:05            & 17.28\%         \\
            5              & 70               & [0.53, 0.05, 0.05] & [12583, 0.00, 0.08] & 5\%            & 0.90   & 0.00 & 59:29            & 10.19\%         \\
            \bottomrule
      \end{tabular}
      \caption{\overtaking results, sorted by the last column. For each combination of number of frequencies $M$ and lattice size (i.e., number of lattice points per dimension), we report the values of $c$, $\lambda$, the runtime, and the achieved lower bound on the safety probability.}
      \label{tab:results-overtaking}
\end{table}

\fi

\end{document}